\documentclass[useAMS,usenatbib]{mn2e}
\usepackage{natbib}
\usepackage{amssymb}
\usepackage{txfonts}
\usepackage{graphicx}
\usepackage{subfigure}
\usepackage{longtable}
\usepackage{verbatim}
\usepackage{color}
\usepackage{longtable}

\def\lesssim{\mathrel{\hbox{\rlap{\hbox{\lower4pt\hbox{$\sim$}}}\hbox{$<$}}}}
\def\gtrsim{\mathrel{\hbox{\rlap{\hbox{\lower4pt\hbox{$\sim$}}}\hbox{$>$}}}}

\newcommand{\Ha}{H$\alpha$}
\newcommand{\Msun}{\mbox{M}$_{\odot}$}
\newcommand{\WHz}{$\mbox{W\,Hz}^{-1}$}

\title[Triggered star-formation in Centaurus A]
{Triggered star-formation in the inner filament of Centaurus A}

\author[R. M. Crockett {\rm et al.}]
{R. Mark Crockett$^1$, Stanislav S. Shabala$^{2,1}$\thanks{Email: Stanislav.Shabala@utas.edu.au}, Sugata Kaviraj$^{1,3}$, Vincenzo Antonuccio-Delogu$^4$, \newauthor Joseph Silk$^1$, Max Mutchler$^5$, Robert W. O'Connell$^6$, Marina Rejkuba$^7$, Bradley C. Whitmore$^5$,\newauthor Rogier A. Windhorst$^8$\\
$^1$Department of Physics, University of Oxford, Keble Road, Oxford, OX1 3RH, UK\\
$^2$School of Mathematics \& Physics, University of Tasmania, Private Bag 37, Hobart, Tasmania 7001, Australia\\
$^3$Blackett Laboratory, Imperial College London, London SW7 2AZ, UK\\
$^4$INAF Ð Osservatorio Astrofisico di Catania, Via S. Sofia 78, Catania I-95123, Italy\\
$^5$Space Telescope Science Institute, Baltimore, MD 21218, USA\\
$^6$Department of Astronomy, University of Virginia, Charlottesville, VA 22904-4325, USA\\
$^7$ESO, Karl-Schwarzschild-Strasse 2, 85748 Garching bei M\"unchen, Germany\\
$^8$School of Earth and Space Exploration, Arizona State University, Tempe, AZ 85287-1404, USA\\
}

\pagerange{\pageref{firstpage}--\pageref{lastpage}} \pubyear{2011}

\begin{document}
\bibliographystyle{aa}
\label{firstpage}

\maketitle

\begin{abstract}

We present recent {\em Hubble Space Telescope} observations of the inner filament of Centaurus~A, using the new Wide Field Camera 3 (WFC3) $F225W, F657N$ and $F814W$ filters. We find a young stellar population near the south-west tip of the filament. Combining the WFC3 dataset with archival Advanced Camera for Surveys (ACS) $F606W$ observations, we are able to constrain the ages of these stars to $\lesssim$10~Myrs, with best-fit ages of 1-4~Myrs. No further recent star-formation is found along the filament.

Based on the location and age of this stellar population, and the fact that there is no radio lobe or jet activity near the star-formation, we propose an updated explanation for the origin of the inner filament. \citeauthor{1993ApJ...414..510S} suggested that radio jet-induced shocks can drive the observed optical line emission. We argue that such shocks can naturally arise due to a weak cocoon-driven bow shock (rather than from the radio jet directly), propagating through the diffuse interstellar medium from a location near the inner northern radio lobe. The shock can overrun a molecular cloud, triggering star-formation in the dense molecular cores. Ablation and shock heating of the diffuse gas then gives rise to the observed optical line and X-ray emission. Deeper X-ray observations should show more diffuse emission along the filament.

\end{abstract}


\section{Introduction}
\label{sec:introduction}

The peculiar elliptical galaxy NGC~5128 is an archetypal example of a post-merger system, believed to have formed via a merger between an early-type galaxy and a gas-rich disk galaxy \citep{1998A&ARv...8..237I}. Located in a poor group at a distance of 3.7~Mpc, it hosts the closest powerful extragalactic radio source, Centaurus~A. These factors make Centaurus~A the perfect laboratory for studies of star-formation and the interaction between the energetic output of supermassive black holes and their surroundings.

\begin{figure*}
\centering
\includegraphics[width=160mm]{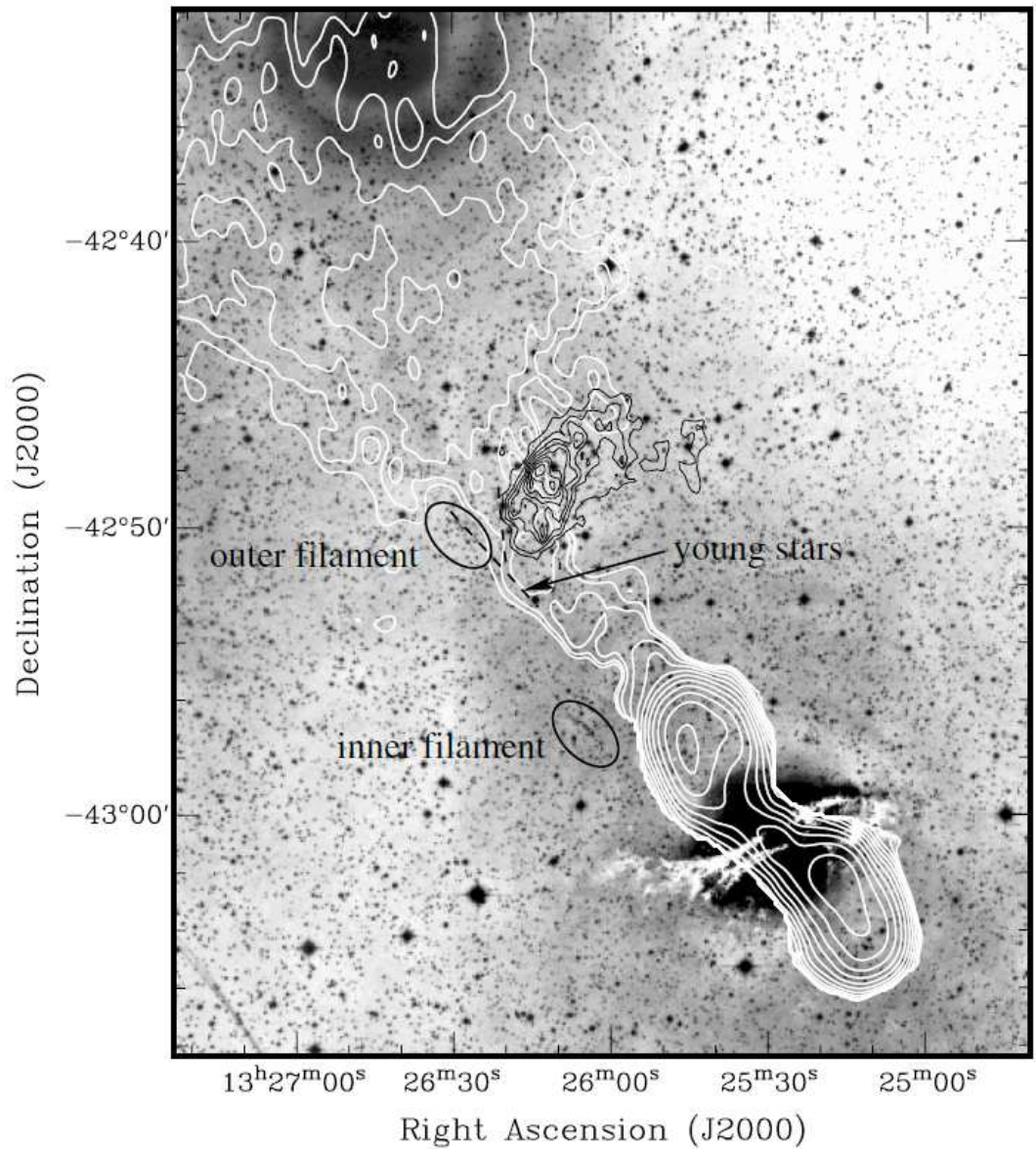}
\caption{Reproduction of Fig.1 from \citet{2005A&A...429..469O}.  Optical image (provided by D. Malin) showing the location of the inner and outer filaments relative to the nucleus of Cen A, and the radio jet (white contours).  The black contours show the position of the HI cloud, associated with the outer filament, which is crossing the radio jet.   Angular size of $1'$ corresponds to 1.07~kpc.  Note that the large diffuse pattern in the top left corner is an image artefact and not a real object.}
\label{fig:CenA_large}
\end{figure*}

Centaurus~A has been extensively studied at wavelengths ranging from radio all the way through to $\gamma$-rays. Radio images \citep[e.g.][]{1999MNRAS.307..750M} show multiple synchrotron-emitting lobes, potentially indicative of restarting activity \citep{2001ApJ...563..103S,1999MNRAS.307..750M}. A pair of inner lobes at 5 kpc from the nucleus co-exist with a single Northern Middle Lobe at 25-30 kpc (Fig.~\ref{fig:CenA_large}). On the largest scales, a pair of outer lobes is seen at around 100 kpc from the nucleus \citep{2009ApJ...707..114F}. The bent nature and re-orientation of the jets between these three scales suggests interactions between the jets and gaseous environment are important \citep{2010NewA...15...96G}. Further evidence for this is provided by observations of X-ray knots along the radio jet, most likely triggered by jet interaction with dense gas clumps \citep{2009ApJ...698.2036K}. 

It is well-known that interaction between radio sources and their environment can result in significant quenching of gas cooling and star-formation. Such feedback is often invoked to explain the lack of significant recent star-formation in massive ellipticals \citep{2006MNRAS.370..645B,2009ApJ...699..525S}. Recently, \citet{2011MNRAS.413.2815S} have shown that radio sources can suppress star-formation on scales of groups and clusters, by heating and sweeping gas out of satellite galaxies. However, radio jets can also trigger star-formation by compressing dense gas with short radiative cooling times. This phenomenon is modelled in simulations \citep[e.g.][]{2009MNRAS.396...61T}, and explains the so-called alignment effect between the radio, UV and optical line emission along the jet axis in distant radio galaxies \citep{1993ARA&A..31..639M,1996MNRAS.280L...9B}.

Two filaments, observed as narrow-line optical emission, are found in the vicinity of the Northern Middle Lobe (Fig.~\ref{fig:CenA_large}) and oriented more or less in the direction of the inner radio jet \citep{1999MNRAS.307..750M,1975ApJ...198L..63B,1975PASAu...2..366P}. The inner filament is located 8.5~kpc from the nucleus, some 2 kpc away from the radio jets \citep{1999MNRAS.307..750M}, while the outer one is 18 kpc from the central engine and aligned with the radio emission.  Broad-band optical observations reveal that both filaments contain hot, young stars of $\lesssim$10 Myrs old \citep{2001A&A...379..781R,2002ApJ...564..688R,2004A&A...415..915R,2000ApJ...538..594F,2002ApJ...575..712G}.  However, while the young stars appear to be evenly distributed along the outer filament \citep{2001A&A...379..781R,2000ApJ...538..594F}, those in the inner filament are concentrated at its south-western tip, closest to the galaxy nucleus \citep{2002ApJ...564..688R}.

H\,I observations highlight another difference between the filaments. The outer filament lies in very close proximity to a large H\,I cloud \citep{1994ApJ...423L.101S,2005A&A...429..469O}. The transverse dimension of the cloud is comparable to the filament length, and the rotational velocity of the cloud supports it from collapsing towards the nucleus \citep{1994ApJ...423L.101S}. Moreover, the stellar age of $\sim 10-15$~Myrs \citep{2002ApJ...564..688R} is consistent with the crossing time of the cloud across the radio jet, and the ionized gas velocities within the filamentary structure are similar to the cloud velocity \citep{1994ApJ...423L.101S,1983ApJ...269..440G,1998ApJ...502..245G}. This scenario led a number of authors \citep[e.g.][]{2002ApJ...564..688R,2010NewA...15...96G} to suggest that the outer filament is a prime example of recent jet-induced star-formation, which is probably still ongoing \citep{2000ApJ...536..266M,2004A&A...415..915R}. On the other hand, no such HI cloud is observed in the vicinity of the inner filament.

Despite the lack of cold gas and distributed stellar populations, the inner filament appears brighter and more tightly collimated in optical emission lines than the outer filament. This emission is also less uniform across the filament, appearing much brighter at the end nearest to the nucleus \citep{1991MNRAS.249...91M}. Two potential mechanisms have been proposed as being responsible for the line emission: photoionization by the nucleus \citep{1991MNRAS.249...91M} and shock heating \citep{1993ApJ...414..510S}. The photoionization model requires Centaurus~A to be a blazar, which is at odds with radio observations of its jets \citep{2003ApJ...593..169H}. It also fails to account for the highest excitation lines and complex velocity structure in the inner filament \citep{2004ApJ...617..209E,1981ApJ...247..813G,1991MNRAS.249...91M}. The jet-driven shock excitation model suggested by \citet{1993ApJ...414..510S} can explain both the high and low-excitation emission by invoking fast and slow-moving shocks. However, this model explicitly requires the presence of a radio jet in the vicinity of the filament, which is not observed. It also fails to predict the distributed X-ray emission found by \citet{2004ApJ...617..209E}.

In this work, we employ high-resolution {\em Hubble Space Telescope} (HST) Wide Field Camera 3 (WFC3) observations to examine star-formation in the inner filament.  The unique spatial resolution and sensitivity of WFC3, in particular at near-ultraviolet (NUV) wavelengths, enables us to isolate and estimate reliable ages for stars and stellar populations.  The location and ages of these stars place new constraints on the possible origin and evolution of the inner filament.

The paper is structured as follows. In Section~\ref{sec:observations} we describe the observations and object detection. Identification of stellar populations and derivation of parameters (in particular, stellar ages) is presented in Section~\ref{sec:sf}. We discuss the implications of our findings in the context of triggered star-formation and the AGN - gas interaction in Section~\ref{sec:interpretation}. We conclude in Section~\ref{sec:summary}.

Throughout the paper, a distance of $3.68 \pm 0.43$~Mpc ($\mu = 27.83 \pm 0.24$) to Centaurus~A is assumed, giving a scale of $1' = 1.07$~kpc.  This distance is taken from NED\footnote{http://nedwww.ipac.caltech.edu/cgi-bin/nDistance?name=NGC+5128} and is the mean of several distance estimates made using a range of methods: Cepheid variables \citep{2007ApJ...654..186F}; Type Ia supernovae \citep[e.g.][]{2007ApJ...659..122J,1992MmSAI..63..233R}; surface brightness fluctuations \citep[e.g.][]{2007ApJ...654..186F,1993AJ....106.1344S}; tip of the red giant branch \citep[e.g.][]{2007ApJ...661..815R,2004A&A...413..903R,1996ApJ...465...79S}; globular cluster radius \citep{2009ApJ...705.1533C}; globular cluster luminosity function \citep{1984ApJ...287..175H}; planetary nebula luminosity function \citep{2002ApJ...577...31C,1993ApJ...414..463H}; Mira variables \citep{2004A&A...413..903R}.

\section{Hubble Space Telescope Observations}
\label{sec:observations}

Our recent HST observations of the Cen~A inner filament were made using the new WFC3 as part of the WFC3 SOC (Scientific Oversight Committee) Early Release Science Program (HST program 11360, PI: R. O'Connell).  These data, taken 2010 Jul 02, were obtained using the $F225W, F657N$ and $F814W$ filters (NUV, H$\alpha$ + [N II], and $I$-band respectively).  Of particular note is the $F225W$ (NUV) data.  Later in this paper, we exploit the sensitivity of the NUV to stellar age in order to age date the very recent star-formation in and around the inner filament. 

Earlier HST observations of the Cen~A inner filament from 2004 Aug 10\&11 were found in the HST archive (HST program 10260, PI: W. Harris).  Taken with the Advanced Camera for Surveys (ACS), these data comprise of three separate pointings, all in the $F606W$ filter.  This very broad $`V'$-band filter spans a number of prominent emission lines which dominate the optical spectrum of the inner-filament -- most notably [OIII] 4959\AA,5007\AA, [OI] 6300\AA,6370\AA, [NII] 6548\AA,6584\AA, \Ha\/, and [SII] 6725\AA.  The filamentary structure is therefore strongly detected in the $F606W$ image.

The ACS data sets were downloaded from the HST archive\footnote{http://archive.stsci.edu/hst} at the Space Telescope Science Institute (STScI) via the on-the-fly recalibration (OTFR) pipeline, and we adopted the pipeline {\em drizzled} data as our science images.  The WFC3 data were bias, dark and flat-field corrected locally using {\sc calwfc3}, and subsequently drizzled using the {\sc multidrizzle} image reconstruction software.  The purpose of {\sc multidrizzle} (both the pipeline and local implementations) was to align exposures in each filter, correct for geometric distortion, remove defects such as cosmic rays and hot pixels, and combine the exposures using the {\em drizzle} image reconstruction technique of \citet{2002PASP..114..144F}.  The latest calibration files, including image distortion coefficient tables (IDCTAB), were downloaded from the WFC3 reference file website\footnote{http://www.stsci.edu/hst/observatory/cdbs/SIfileInfo/WFC3/reftablequeryindex}.  Further details of WFC3/UVIS data reduction are given by \citet{2011ApJS..193...27W}.

\begin{table}
\caption{HST observations of the Cen A inner filament}
\begin{center}
\begin{tabular}{lllc}
\hline\hline
Date & Instrument & Filter & Exposure Time\\
& & & (s) \\
 \hline
2004 Aug 10 & ACS/WFC&F606W & 2370\\
2004 Aug 10 & ACS/WFC &F606W & 2370\\
2010 Jul 02 & WFC3/UVIS &F225W & 2232\\
2010 Jul 02 & WFC3/UVIS &F657N (H$\alpha$ + [N II]) & 1640\\
2010 Jul 02 & WFC3/UVIS & F814W & 1050\\
\hline\hline
\end{tabular}
\end{center}
\label{tab:obs_table}
\end{table}

The Cen~A inner filament is strongly detected in ACS $F606W$ and WFC3 $F657N$ as optical line emission from ionized gas.  It is situated $\sim$ 8.5 kpc to the north-east of the galaxy nucleus, and extends a further 2200 pc along the same direction, with a width of $\sim$50-100 pc. (Note all of the above are projected lengths).  The first 500 pc of the filament, beginning in the south-west closest to the galaxy nucleus, consists of three major groups of extended emission.  These appear as unresolved `blobs' in the seeing-limited observations of \citet{1991MNRAS.249...91M}(see their Fig. 2a), who label the regions A, B and C (following \citealt{1975ApJ...198L..63B,1978ApJ...226L..79O}).  The HST images in Figs.~\ref{fig:inner_filament}\,--\,\ref{fig:colour_ABC} reveal significant structure in these regions. {\em Shell-like} features are visible, which presumably demarcate regions of higher density gas (perhaps due to compression by shocks), and appear to lie perpendicular to the line extending from the galaxy nucleus, and along the inner filament.

The filament continues to the north-east with decreased surface brightness, and appears to be split into two further regions, labelled E and F by \citet{1991MNRAS.249...91M}.\footnote{The lack of a region `D' in \citet{1991MNRAS.249...91M} relates to the labeling scheme of \citet{1975ApJ...198L..63B}, who identified regions A, B and C of the inner filament, and used `D' to denote what is now referred to as the outer filament.}  The morphologies of regions E and F are less well defined in the HST imagery due to low surface brightness.  However, there are tentative traces of shell structures orthogonal to the main axis of the filament, similar to those seen in regions A to C.

\subsection{Object detection}
\label{sec:object_detection}

A unique aspect of the WFC3 dataset is the inclusion of $F225W$ observations.  These diffraction limited NUV images are most sensitive to the hottest, youngest stars, and offer us clear information as to their properties and location in and around the inner filament.  Given that other purely optical studies have already been published \citep[e.g.][]{2002ApJ...564..688R} we have focussed our attention on those objects that are detected in $F225W$.

We used {\sc daofind} -- part of the {\sc iraf daophot} package \citep{1987PASP...99..191S} -- to automatically identify point sources in the $F225W, F606W$ and $F814W$ data.  A 5$\sigma$ detection threshold was applied.  The resulting coordinate tables were cross-correlated to extract those objects detected in the $F225W$ image and at least one other filter.  Cross correlation of the ACS and WFC3 object positions required a transformation between the ACS $F606W$ and WFC3 $F814W$ coordinate grids, which was calculated using the {\sc iraf} task {\sc geomap} and the positions of 37 point sources common to both images.  The ACS $F606W$ coordinates were then transformed to the WFC3 coordinate frame using the task {\sc geoxytran}.

A final visual inspection of the data was carried out to check for any other common sources missed by {\sc daofind} and cross-correlation, adding 17 extra objects to the list.  Several of these were saturated objects that had been rejected by {\sc daofind} due to their shape characteristics.  Other added objects were just below the detection threshold and/or partially blended with neighbouring bright objects in one or more filters.

The positions of all objects detected in $F225W$ and at least one other broadband filter are plotted on the two images in Fig.~\ref{fig:NUV_object_posns}.  The position markers have been colour-coded in relation to the observed colour of each of the objects, and this photometry will be described fully in the following section.  However, it is useful to discuss now the relative morphologies of the two main populations we see in this figure -- those with blue NUV-optical colours [($F225W-F814W$) $<$ 0; blue markers] and those with red colours [($F225W-F814W$) $>$ 0; red and green markers].

The most obvious difference is in their distributions across the field-of-view.  Red objects appear quite evenly spread, while blue objects are mostly to the right (south west) of the frame, closest to the galaxy nucleus and the radio jet.  

Many of the red objects are very bright, or even saturated, in $F606W$ and $F814W$ and have diffraction spikes or show clear diffraction patterns around the central peak of their PSFs.  We conclude that these particular objects are most likely foreground stars in our own Milky Way.  

The highest concentration of blue objects is found at the SW-tip of the inner filament, which also harbours the brightest of the blue objects.  It is also the only region that such blue objects are found in association with strong line-emission.  The colours and magnitudes of these objects (discussed more fully in the following sections) coupled with the line-emission suggests the presence of young stars or star clusters.  Figs.~\ref{fig:3_filters_close},\,\ref{fig:colour_ABC}\,\&\,\ref{fig:colour_regions} show progressively magnified views of the SW-tip in greyscale and in colour, where the RGB colour images have been constructed by assigning $F814W$, $F606W$, and $F225W$ to the red, green and blue channels respectively.  The hot, blue stars are clearly seen concentrated in two regions at the SW-tip.  No other blue stars are seen anywhere else along the emission-line filament \footnote{5$\sigma$ detection limits (Vegamag) are: $F225W$ = 25.0; $F606W$ = 27.4 (no emission-line background), $F606W$ = 26.8 (strong emission-line background); $F814W$ = 25.8}. 

The reader should note that the speckled background visible in the $F606W$ and $F814W$ images in Figs.~\ref{fig:NUV_object_posns}\,\&\,\ref{fig:3_filters_close} or in the colour images shown in Figs.~\ref{fig:colour_ABC}\,\&\,\ref{fig:colour_regions}) is not due to noise, but rather is the myriad of background stars detected in the body of Cen A.  Interestingly, the stellar field coincident with the emission-line filament is indistinguishable from this general background, except at the SW-tip where we see the dense concentration of (presumably) young, NUV-bright stars.  This suggests that recent star-formation in the inner filament has been predominantly confined to the SW-tip, and that no -- or at least no significant -- recent star-formation has taken place anywhere else along the filament.

This contradicts the observations of \citet{2002ApJ...564..688R} who reported the detection of young stars ($\lesssim$ 10 Myrs) along the entire length of the inner filament (although they also noted the concentration at the SW tip).  Our Fig.~\ref{fig:rejkuba_overlay} is a partial reproduction of Fig. 4 from \citet{2002ApJ...564..688R}, showing the positions of the bluest (U-V) $<$ -0.5 stars (black triangles) which they detected along the inner filament.  We have overlaid a colour composite image, constructed from the same broadband $U$ (blue), $V$ (green) and $I$ (red) Magellan Telescope observations as used by \citeauthor{2002ApJ...564..688R} in their analysis.  It is clear that the majority of these apparently blue stars lie along the filament, which also appears blue in this figure, most likely due to [OII] 3727$\AA$ line-emission detected in the broad $U$-band filter.  Apart from the SW tip and a few other outlying sources, none of these blue stars are detected in our WFC3 $F225W$ image.  Given the resolution and depth of the HST data, we would have expected strong NUV detections if hot, young stars were present.

If the dust extinction along the rest of the filament is much higher than at the SW tip, it might be possible that the NUV light from the young stars is fully attenuated.  However, \citet{2002ApJ...564..688R} find no evidence for such high extinction.  Another explanation might be that the $U$-band photometry from  \citeauthor{2002ApJ...564..688R} is contaminated by [OII] 3727$\AA$ line emission, leading to artificially bluer photometry.  Lacking narrow-band imagery, \citeauthor{2002ApJ...564..688R} had to rely on {\em sky-background} subtraction to correct for underlying line-emission in their stellar photometry, particularly in the $U$-band.  We can see from the Magellan and HST images presented here that the surface brightness of the filament changes rapidly on short spatial scales.  This makes it difficult to determine an appropriate sky-background level for any individual object along the filament.  For example, where an object happens to coincide with a region of higher emission-line flux, the background level determined from the surrounding pixels will be an underestimate, leading to brighter source photometry.  

This may have been the case with some of the $U$-band photometry in \citet{2002ApJ...564..688R}, and would naturally explain why the majority of their blue star candidates are found associated with the brightest regions of the filament.  Of course, one could also argue that the association of young stars with regions of brighter line-emission is a direct consequence of their photoionization of the surrounding gas.  Nevertheless, without narrow-band data it is difficult to rule out contamination in the $U$-band photometry, and our WFC3 $F225W$ observation strongly suggests that {\em very} young stars are almost entirely confined to the SW tip.

In light of our HST data, it seems reasonable to suggest that photoionization by young stars ($<$10 Myr old) is contributing to the strong line-emission at the SW tip.  Indeed, \citet{1981ApJ...247..813G} have already shown that a spectrum of this region is similar to that of a normal HII region powered by embedded, hot stars.  However, the apparent lack of young stars along the rest of the filament, and the appearance of high excitation lines, suggests that some other mechanism must be exciting the gas in these regions.  As mentioned in the introduction, two mechanisms have already been proposed to explain the line emission in the inner filament: photoionization by the galaxy nucleus \citep{1991MNRAS.249...91M} and shock heating \citep{1993ApJ...414..510S}.  Later in this paper we present a model, based on the Sutherland et al. shock heating hypothesis, which incorporates the formation of the inner filament, the X-ray flux and optical line emission, and the recent star-formation at the south-western tip.  In the following section we focus on the star-formation, and attempt to fit ages to the young stellar populations using HST photometry.


\section{Star-formation in the inner filament}
\label{sec:sf}

\begin{figure*}
\centering
\includegraphics[width=160mm]{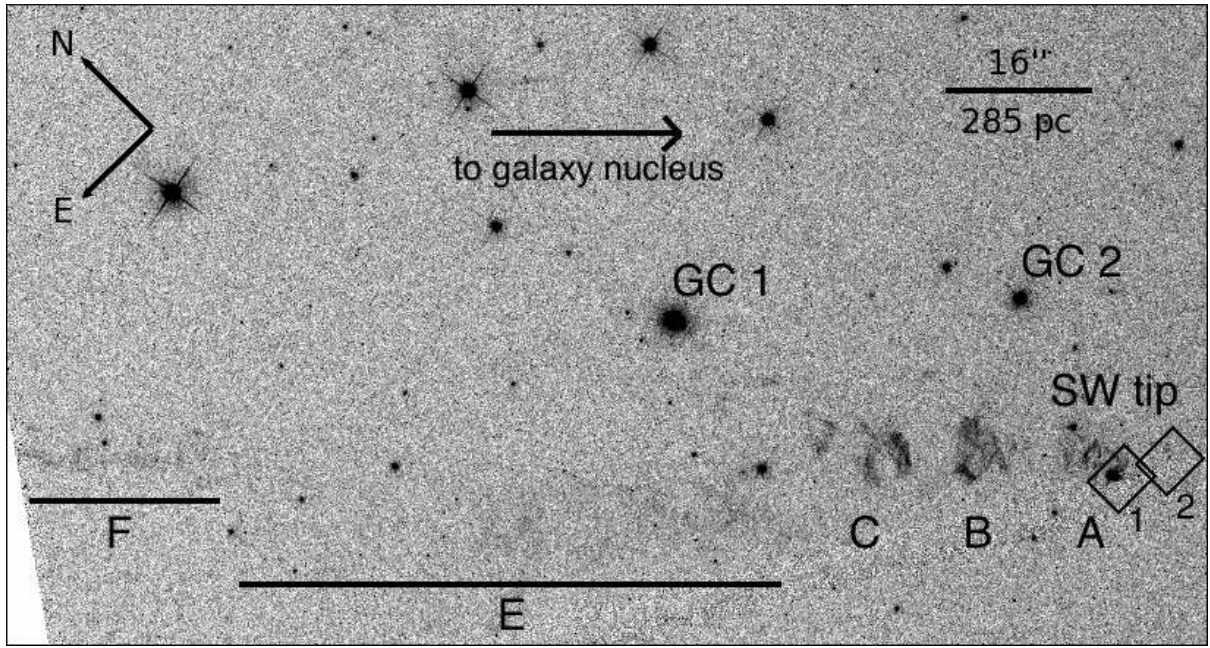}
\caption{Combined WFC3 $F657N$ (\Ha\/ + [NII]) and $F814W$ image of the Cen A inner filament and its immediate environment.  The filament, observed here as \Ha\/ emission, is split into five major regions labelled A, B, C, E and F, following the labelling scheme of \citet{1991MNRAS.249...91M}.  (The absence of a region D is historical and relates to the discovery of regions A, B and C of the inner filament by \citet{1975ApJ...198L..63B}, who denoted the outer filament as region D.)  The filament has a projected length of $\sim125\arcsec$ (2200 pc), and width of $\sim3-6\arcsec$ (50-100pc).  Its appears, at least in projection, to lie along a line originating at the galaxy nucleus and directed to the north-east.  The majority of bright sources to the north-west of the filament are Galactic stars along the line of sight, although two Cen A globular clusters are marked (GCs 1 \& 2).  No young stars are found along the length of the filament, except in two regions in the south-western tip (labelled regions 1 \& 2).}
\label{fig:inner_filament}
\end{figure*}

\begin{figure*}
\centering
\includegraphics[width=170mm]{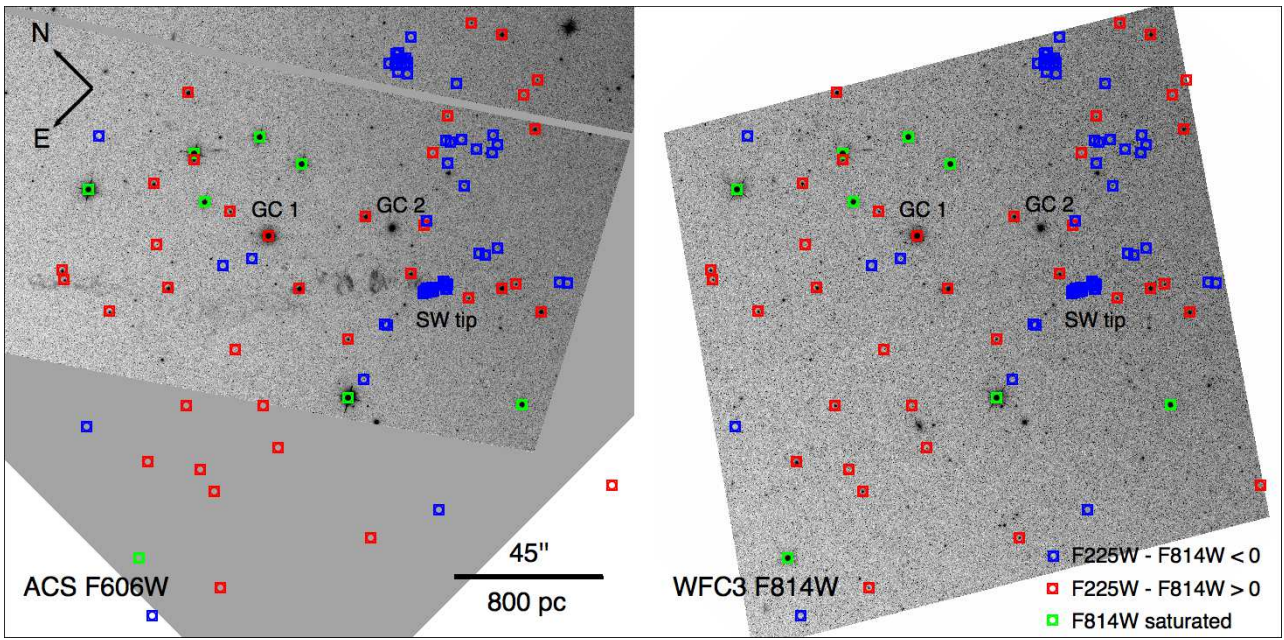}
\caption{ACS F606W and WFC3 F814W images showing the locations of all objects detected in F225W and at least one other filter.  Blue markers indicate objects with observed ($F225W-F814W$) $<$ 0, while red markers denote those with ($F225W-F814W$) $>$ 0.  Green markers denote objects that are saturated in the WFC3 F814W data.  The different pointing of the ACS F606W image results in several objects falling outside the field-of-view to the south east.  The inner filament is clearly visible in the centre of the F606W frame, running right to left (south west to north east).  A concentration of blue, NUV-bright objects is located at its south western tip, but no other such objects are visible along the rest of the filament.  Other blue objects are detected in the rest of the field -- although nowhere with the same concentration as observed at the SW-tip -- with the vast majority located in the west between the filament and the radio jet.  The red and saturated objects (red and green markers) are more evenly distributed across the field-of-view, and many of these are likely foreground stars in our own Galaxy.  Some, including GC 1, are consistent with being globular clusters in Cen A.}
\label{fig:NUV_object_posns}
\end{figure*}

\begin{figure*}
\centering
\includegraphics[width=160mm]{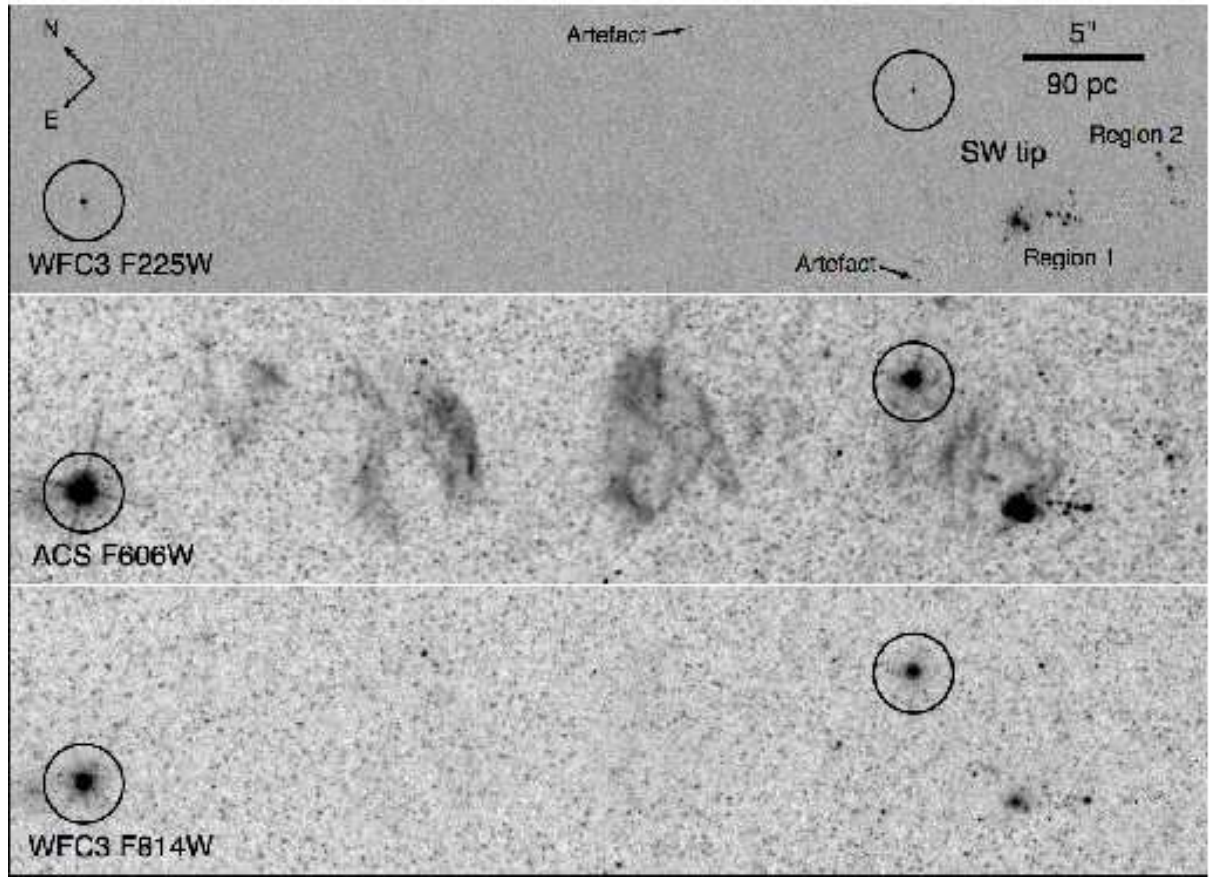}
\caption{WFC3 F225W, ACS F606W and WFC3 F814W image sections showing magnified view of regions A, B and C of the inner filament, including the SW-tip.  Black circles mark the position of two likely foreground stars.  It is immediately clear from the F225W image that the NUV-bright sources in this figure are confined entirely to the two regions in the SW-tip.  No other NUV point-sources are detected coincident with any of the line emission anywhere else along this section of the filament.}
\label{fig:3_filters_close}
\end{figure*}

\begin{figure*}
\centering
\includegraphics[width=160mm]{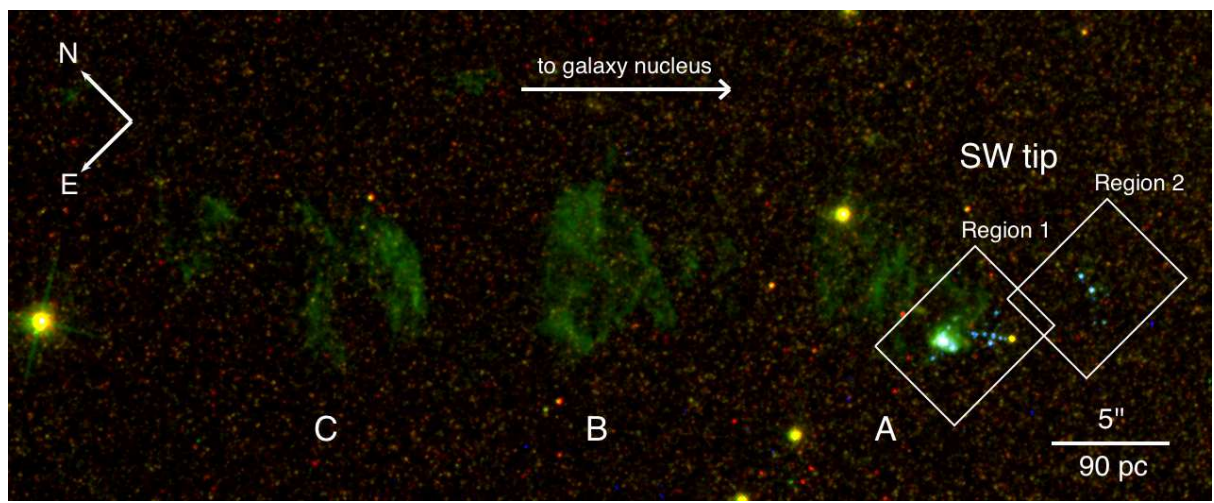}
\caption{RGB colour composite image of regions A, B and C in the Cen A inner filament.  This image was created using the WFC3 $F225W$ (blue), ACS $F606W$ (green) and WFC3 $F814W$ (red) observations.  The \Ha\/ emission of the filament falls within the $F606W$ passband and is therefore visible in this figure as diffuse green emission.  At HST resolution, regions A, B and C reveal a {\em shell-like} structure, with the shells oriented perpendicular to the main axis of the filament.  No young stars are observed coincident with the filament, except at the south-west tip in the regions marked 1 and 2.  In these fields we see significant NUV-flux, consistent with the presence of young, massive stars.}
\label{fig:colour_ABC}
\end{figure*}

\begin{figure*}
\centering
\begin{tabular}{cc}
\vspace{2mm}
\includegraphics[width=70mm]{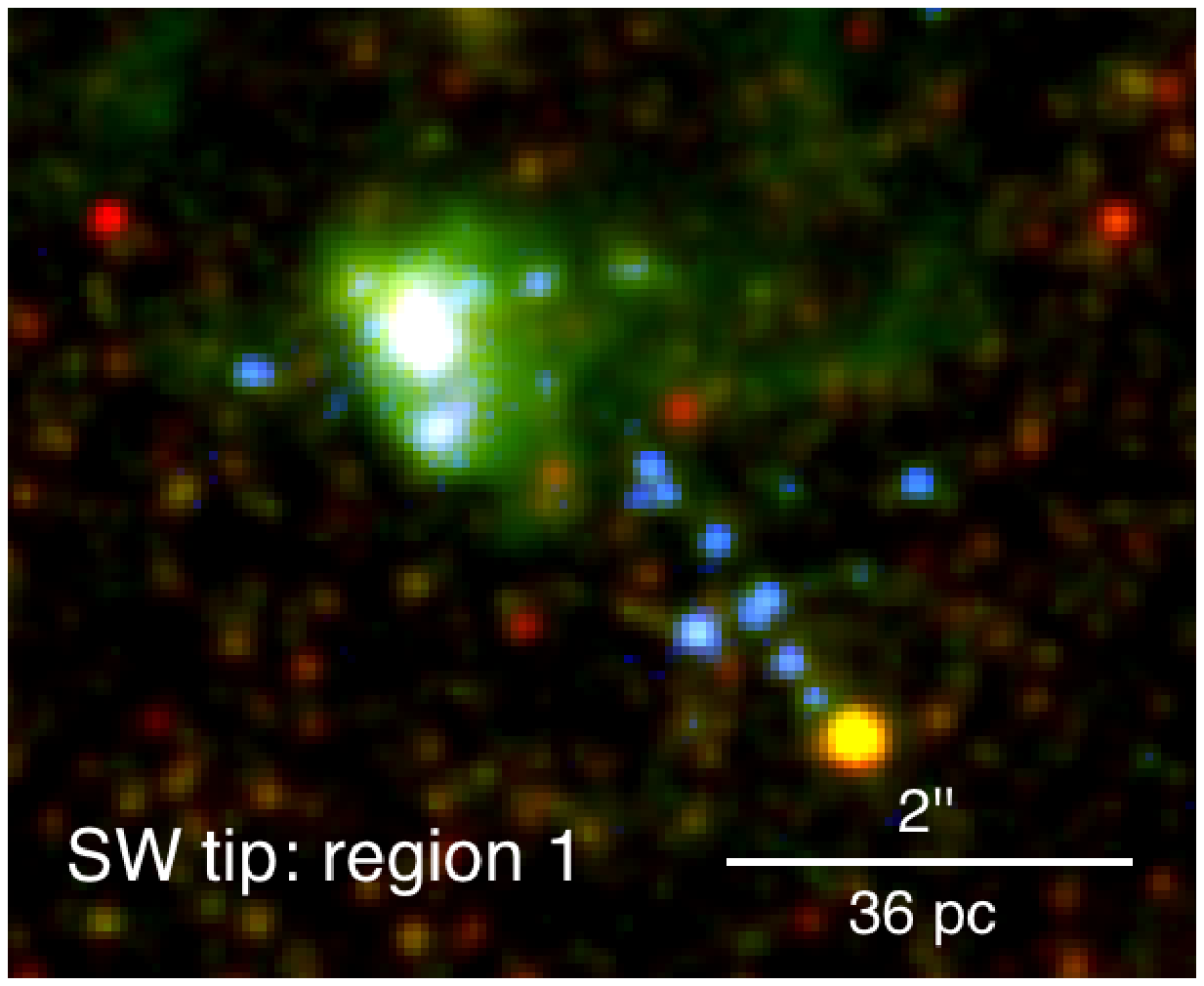}&
\includegraphics[width=70mm]{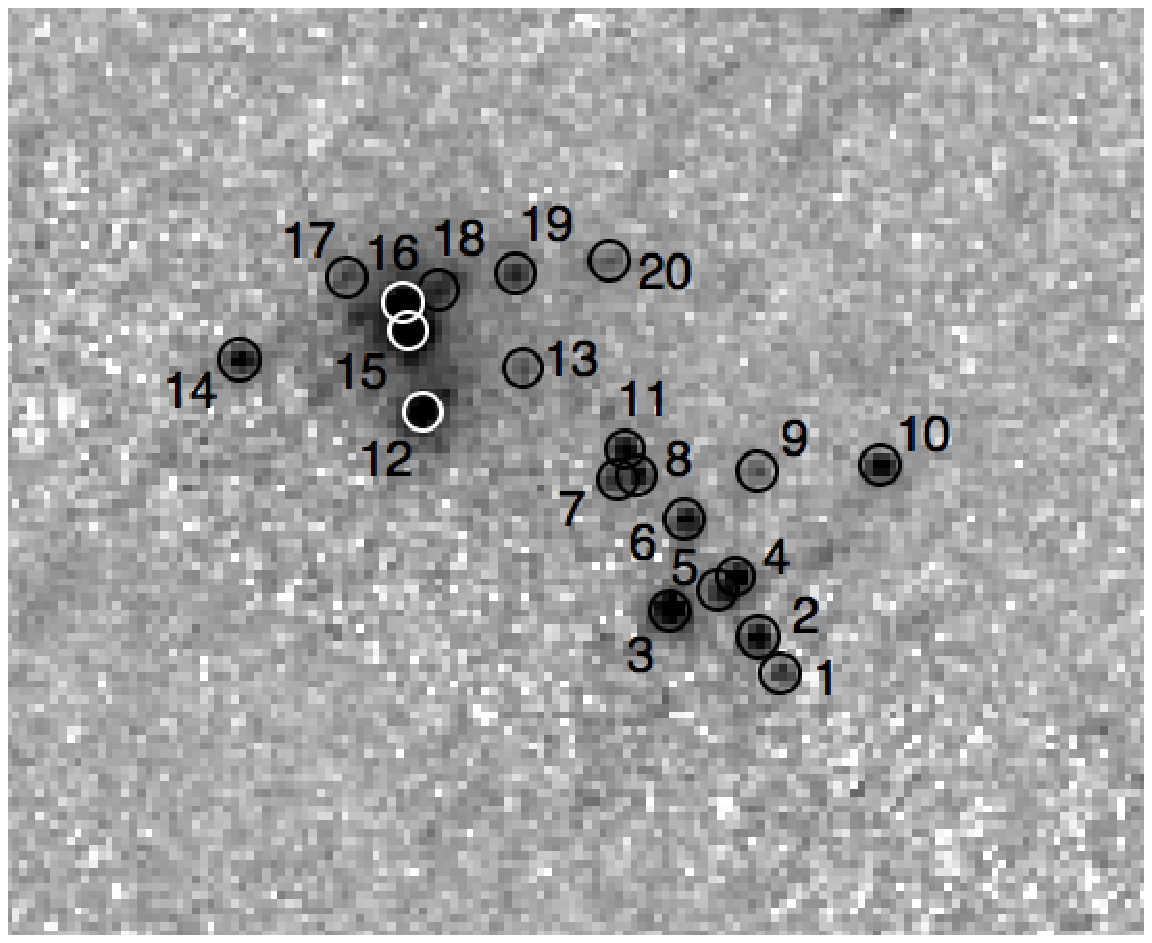}\\
\includegraphics[width=70mm]{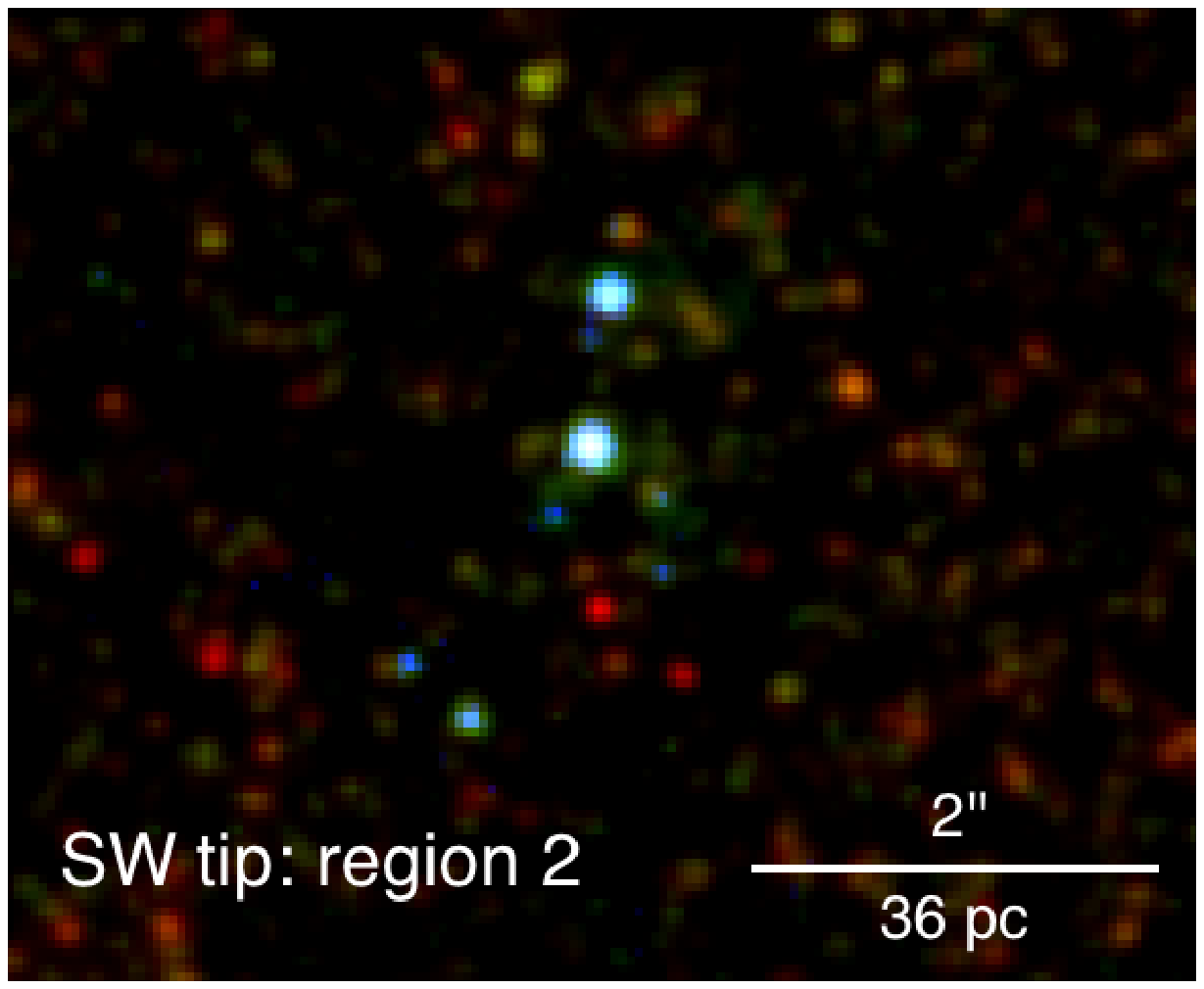}&
\includegraphics[width=70mm]{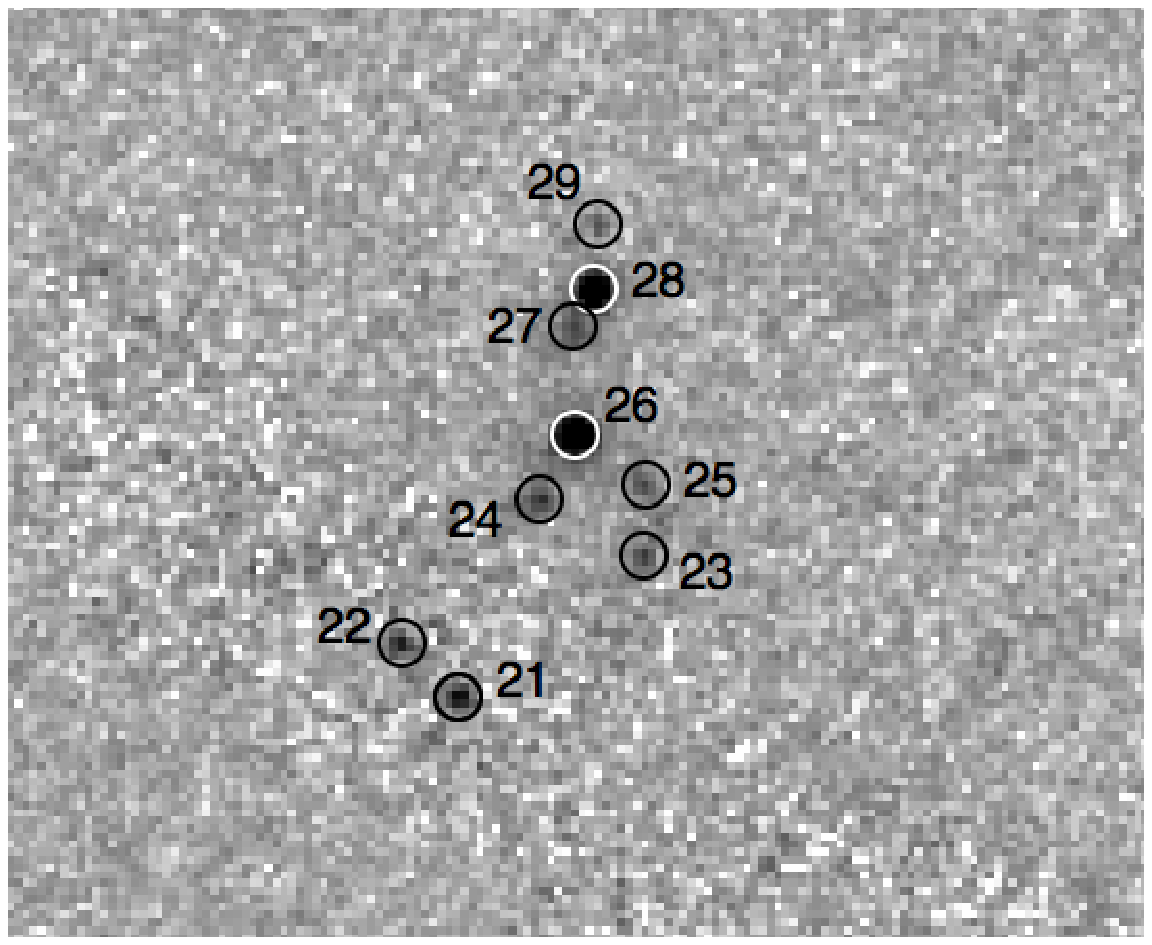}\\
\end{tabular}
\caption{Magnified colour composite (left) and greyscale $F225W$ (right) images of Regions 1 and 2 in the south-western tip of the Cen A inner filament.  All images are oriented such that north is up and east is to the left.  NUV bright sources are identified in the right-hand panels. The ID numbers correspond to those used in Figures~\ref{fig:region1_pdfs}\,\&\,\ref{fig:region2_pdfs} (see Section~\ref{sec:single_star_fitting}). The reader should note that object 1 is not coincident with the bright yellow object visible in the colour image -- it is the fainter blue object immediately to the north-east.}
\label{fig:colour_regions}
\end{figure*}

\begin{figure}
\centering
\includegraphics[width=80mm]{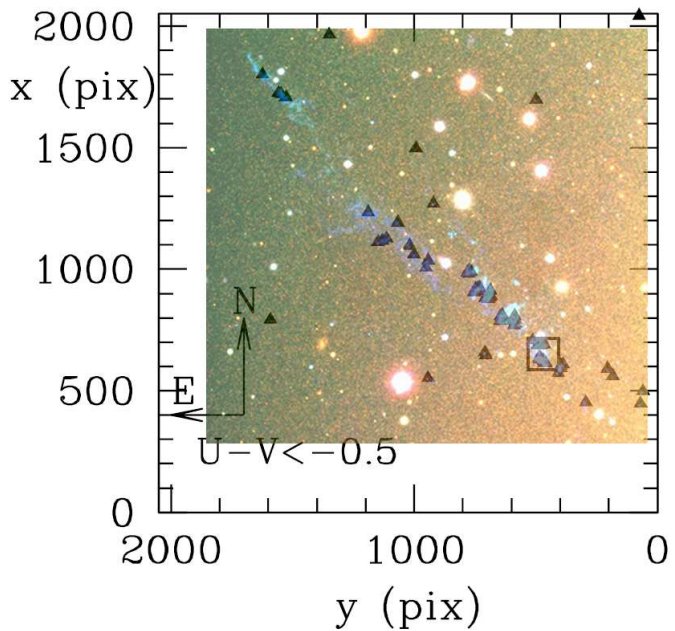}
\caption{Partial reproduction of Fig. 4 of \citet{2002ApJ...564..688R} showing the positions of the bluest (U-V) $<$ --0.5 stars (black triangles) which they detected along the inner filament.  Here, we have overlaid a colour composite image, constructed using the $U$ (blue), $V$ (green) and $I$ (red) Magellan Telescope observations used by \citealt{2002ApJ...564..688R} in their analysis. The small square delineates Knot A of \citet{2002ApJ...564..688R}, which we identify with the SW tip.}
\label{fig:rejkuba_overlay}
\end{figure}

As described above, two regions in the south-western tip of the inner filament (Figs.~\ref{fig:inner_filament}\,\&\,\ref{fig:colour_ABC}) shows signs of young, massive stars; namely bright, compact objects visible in all of the HST observations (NUV to $I$-band).  Fig.~\ref{fig:colour_regions} shows magnified colour composite and greyscale $F225W$ images of these fields.  In the colour images, the $F225W$ data is shown in blue, $F606W$ in green, and $F814W$ in red.

The diffuse green glow prominent in the north of Region 1, and concentrated around the NUV-bright star clusters, is strong line-emission detected in the broad $F606W$ filter.  In contrast, the blue objects in the south of Region 1, in Region 2, and throughout the rest of the field show little or no sign of such line-emission.  This could suggest one or a combination of the following: (1) that the photo-ionizing flux emitted by these stars is too low; (2) that the gas density in these areas is lower than in the north of Region 1; or  (3) that gas has been swept out by stellar winds or shocks.  

Region 2 (situated to the south-west of Region 1 and closer to the galaxy nucleus) contains fewer NUV-bright objects, and shows only traces of line-emission around its two brightest NUV components.  It is detected in [OIII] by \citet{1991MNRAS.249...91M} (see their Fig. 2a), but at much lower intensity than Region 1.  This again suggests lower photo-ionizing fluxes and/or gas densities than in the northern part of Region 1.

Also visible in the colour images is a multitude of yellow and red objects that are not detected in the NUV.  These are apparently part of the background stellar population in Cen A as these stars appear throughout the HST field. From \citet{2004A&A...415..915R}, we expect the red giant branch (RGB) in Cen A to appear at magnitudes $V >$ 25.5 -- 26 mag.  The $F606W$ photometry of the reddish objects is consistent with their being RGB stars, indicating ages of $>$1 Gyr.  It is worth stressing that {\em none} of the red objects visible in Fig.~\ref{fig:colour_regions} closest to the SW-tip are likely supergiants, as their absolute magnitudes (typically M$_{F606W}$ $>$ --2) are too faint.

It is instructive to determine the ages of the NUV-bright stellar populations within the inner filament, particularly in the context of the \citet{1993ApJ...414..510S} shock heating framework.  The possible association of an AGN related shock with a gas cloud and young stars points towards jet-induced star-formation -- i.e. positive AGN feedback.  In the following sections we carry out photometry on the HST ACS and WFC3 datasets, and subsequently compare the observed measurements to models of single stars and stellar populations to determine ages.  Although we present photometry for all sources detected in the $F225W$ filter, our analysis and subsequent discussion is focussed on the dense concentration of blue objects at the SW-tip and the potential link between their formation and that of the emission-line filament.

\subsection{Photometry}

\subsubsection{PSF-fitting}
\label{sec:psf_fitting}

Photometry was carried out using point spread function (PSF) fitting tasks within the {\sc iraf daophot} package \citep{1987PASP...99..191S}.  Vegamag zeropoints for both the WFC3 and ACS instruments were taken from their respective webpages hosted by STScI\footnote{http://www.stsci.edu/hst/wfc3/phot\_zp\_lbn}$^{,}$\footnote{http://www.stsci.edu/hst/acs/analysis/zeropoints}.  Only those sources detected in $F225W$ and at least one other filter were analyzed, and this photometry is presented in Table~\ref{tab:phot_table}.  The matched coordinate lists (as described in $\S$\ref{sec:object_detection}) were input to the {\sc allstar} task, along with model PSFs for each of the broadband filters.  Empirical PSF models were constructed using several, relatively bright stars in each image.  Ideally, isolated stars should be used and this was possible in the $F225W$ image.  The highly crowded $F606W$ and $F814W$ images required a multi-stage process, in which an initial PSF model was created and subsequently used to subtract all objects close to the PSF stars.  A second model was then generated from the {\em cleaned} image and the process repeated until the photometry of the subtracted objects converged.

The crowded nature of the $F606W$ and $F814W$ images in general, and of Regions 1 \& 2 in all filters, necessitated the use of small fitting radii in order to properly measure photometry for close or partially blended objects.  These radii were 1.8 pixels in WFC3/UVIS and 1.5 pixels in ACS/WFC.  Median sky background levels were measured in concentric annuli with internal and external radii of 1.5$\arcsec$ and 2$\arcsec$ respectively.

Contamination of the ACS $F606W$ image by strong line emission posed a particular problem for the photometry of some objects in Regions 1.  Crowding and the rapid variation of the line emission on small spatial scales made it impossible to accurately measure the line emission using annular apertures.  As mentioned previously in $\S$\ref{sec:observations}, the F606W filter spans several prominent emission lines which dominate the optical spectrum of the inner filament.  The strongest of these are [OIII] 5007\AA, and \Ha\/ 6563\AA.

We have used the WFC3 $F657N$ narrowband data to correct the $F606W$ image for \Ha\/ and [NII] 6548\AA,6584\AA emission.  First, we aligned and re-sampled the $F657N$ data to the coordinate grid of the F606W image using the {\sc iraf} tasks {\sc geomap} and {\sc gregister}.  The ratio between the $F606W$ and $F657N$ continuum fluxes was calculated using {\sc synphot} -- part of the {\sc iraf stsdas} package -- and the spectrum of a K-type main sequence star.  Such stars have only very weak \Ha\/ absorption lines and we could therefore assume that the calculated fluxes in both filters represented stellar continuum only.  Appropriate scaling of the $F606W$ data therefore allowed us to approximate the continuum emission of all stellar sources in the $F657N$ image.  Here, we derived a scale factor of 0.04.  The $F606W$ image was scaled and subtracted from the narrowband data resulting in an $F657N$ image free from stellar continuum.  This {\em emission line} image was rescaled by a factor of 1.68 (again, calculated using {\sc synphot}) to take into account the higher throughput of the $F606W$ filter at the \Ha\/ and [NII] wavelengths, and finally subtracted from the original $F606W$ image.  The final $F606W$ image was thereby corrected for contaminating \Ha\/ and [NII] line emission.

Our {\em corrected} $F606W$ image still contains emission from several other lines, but without additional narrowband observations -- in particular $F502N$ which targets the [OIII] 5007\AA line -- it is difficult to remove all of the contamination.  In an attempt to estimate the flux contributions of the remaining emission lines, we created a synthetic spectrum using the line ratios recorded by \citet{1991MNRAS.249...91M} for the area closest to our Region 1.  {\sc synphot} was used to calculate the count rates in the $F606W$ and $F657N$ filters for this synthetic spectrum, from which we estimated that the emission line counts in the broadband image should be around 4.2 times higher than in the narrowband.  However, this high scale factor lead to an {\em over-subtraction} in the vicinity of the brightest stars/star clusters in Region 1, observed as negative count values in the pixels surrounding these objects.  This suggests that the line ratios in the emitting gas coincident with these stars/star clusters might be different from those measured elsewhere in the filament, and indeed varying line ratios are observed by \citet{1991MNRAS.249...91M} -- e.g. [OIII] 5007\AA / \Ha\/.

Since no reliable {\em full}-correction was deemed possible without further observations we have used the $F606W$ image corrected for \Ha\/ and [NII] emission only.  As a result object photometry in areas of strong line emission -- the north of Region 1 (Fig.~\ref{fig:colour_regions}) -- will most likely be artificially brightened due to uncorrected flux.

\subsubsection{Partially resolved objects}

While the majority of objects are relatively well fit by the stellar PSF, a few possess profiles that are significantly broader than that of a point source.  In the case of objects 81 (also labelled GC 1 in Figs.~\ref{fig:inner_filament}\,\&\,\ref{fig:NUV_object_posns}) and 109 (see Table~\ref{tab:phot_table}) the partially resolved appearance is obvious to the naked eye, and they are identified as globular cluster candidates within Cen A.  Photometry was measured in large apertures with radii of 0.80$\arcsec$ and 0.47$\arcsec$ for objects 82 and 109 respectively, the aperture sizes chosen to be roughly 3 times the FWHM of each object in the $F606W$ image.  These radii reflect an approximate cut-off between each cluster and the stellar background. The high density of background point sources made it impossible to reliably isolate objects 82 and 109 in $F606W$ and $F814W$, which subsequently prevented the accurate determination of {\em total} brightnesses.  However, the measured colours are considered reliable.

Objects 12, 15 and 16 -- located in the the northern part of Region 1 (see Fig.~\ref{fig:colour_regions}) -- were also found to be poorly fit by the stellar PSF, suggesting they are likely star clusters.  We employed a modified version of {\sc ishape} \citep{1999A&AS..139..393L} to model each as a convolution of the stellar PSF and an intrinsic source function of variable size and shape.  

The standard implementation of {\sc ishape} does not function well when targets are partially blended, which was the case for objects 15 and 16.  Despite a pixel weighting procedure that was designed to mask-out neighbouring objects, the fitted profiles for these objects invariably fell between their respective centroids.  Our crude {\em modification} was to fix the centroid of the fitted profile at a user-defined position, and to limit the radius within which $\chi^2$-minimization occurs to one smaller than the radius used to extract the fitted model -- analogous to the {\em fitrad} and {\em psfrad} parameters in {\sc daophot} PSF-fitting tasks.  The residual images for objects 15 and 16 from our modified code showed a marked improvement over those produced by the standard version of {\sc ishape}.  

The {\sc ishape} models were used to define filter-dependent aperture corrections, which were subsequently applied to small aperture photometry of objects 12, 15 and 16.  Radii of 1.8 WFC3/UVIS pixels and 1.5 ACS/WFC pixels were chosen to prevent overlapping of apertures during photometry of objects 15 and 16.

\begin{figure*}
\centering
\includegraphics[width=50mm,angle=90]{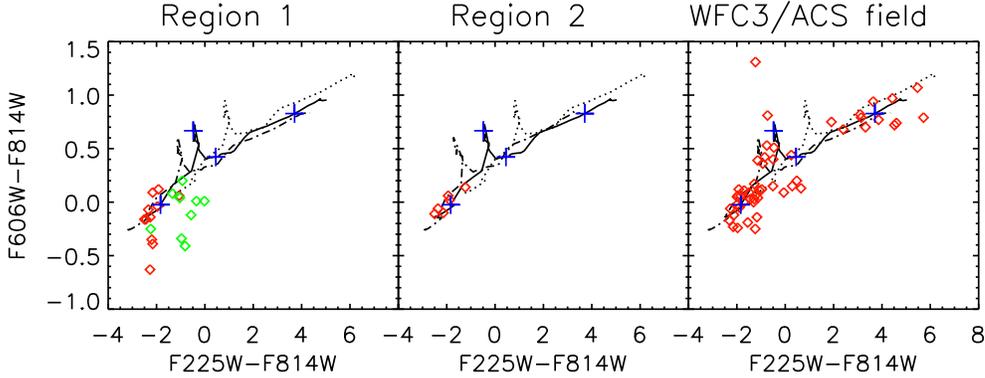}
\caption{Photometry of detected NUV-bright objects (red diamonds) overlaid on model SSPs.  The metallicity of each model grid is solar.  Three extinction levels are shown:  zero (dot-dashed line), foreground (solid line) and 3x foreground (dotted line).  The blue crosses in each panel, going from left to right, indicate the locations of 5, 10, 100 and 1000 Myr populations on the foreground-extinguised model grid.  The photometry of objects in Region 1 has been split into two groups: northern objects associated with strong line-emission (green diamonds); southern objects associated with little or no line-emission (red diamonds).  The northern objects appear significantly redder than those in the south, most probably due to higher extinction.  See text for detailed discussion.}
\label{fig:col_col_SSP}
\end{figure*}

\subsection{Age estimation}
\label{sec:Age_est}

We have taken the conservative approach of comparing {\em all} of the photometry to models of both stellar populations and single stars.  Profile fitting suggested that most of the objects were consistent with the stellar PSF, but one cannot rule out the possibility that some of these might be close multiple systems (binaries etc.) or low-mass, compact stellar clusters based on profile characteristics alone.  While it is true that neither model is particularly suited to these specific possibilities, the comparison of photometry to both models allowed us to identify those objects that are more consistent with one or the other, those that are consistent with both, and those that are not well fitted by either model.

Ulitimately, our main aim was to determine reliable and robust age estimates for the NUV-bright objects in the SW tip of the inner filament.  This was achieved through the convergence of the results from stellar population and single star model fitting as described in the following paragraphs.

\subsubsection{Simple stellar populations}
\label{sec:SSP}

In Fig.~\ref{fig:col_col_SSP} we plot our HST photometry in colour-vs-colour diagrams along with model colours of simple stellar populations (SSPs).  Synthetic WFC3 and ACS photometry for a grid of SSPs based on the Padova stellar models \citep{2000A&AS..141..371G,2008A&A...482..883M} was obtained from the CMD web-interface, hosted by the Osservatorio Astronomico di Padova\footnote{http://stev.oapd.inaf.it/cgi-bin/cmd}.  Model photometry was calculated for ages of $10^6$ to $2\times10^9$ yrs with increments of $\Delta$log(Age[yr]) = 0.05, metallicities of Z = 0.002 to 0.030 (0.1 to 1.5 Z$_{\odot}$) with increments of $\Delta$Z = 0.002, and assuming a \citet{2001MNRAS.322..231K} initial mass function (IMF).  The uncertainties in the synthetic photometry were taken to be 0.05 mags in the optical filters, and 0.1 mags for the NUV \citep{2003ApJ...582..202Y}.  The SSP models shown plotted in Fig.~\ref{fig:col_col_SSP} are of roughly solar metallcity (Z = 0.020) with extinctions of E$(B-V)$ = 0.0, 0.115 and 0.345 \citep[zero, 1$\times$ and 3$\times$ foreground extinction;][]{1998ApJ...500..525S}.

The photometry in Fig.~\ref{fig:col_col_SSP} is shown divided into three groups; Regions 1 and 2, both situated at the SW tip of the inner filament, and the rest of the WFC3/ACS field (see also Table~\ref{tab:phot_table}).  By definition only those objects with photometry in all three broadband filters are plotted in these diagrams, thereby excluding objects that were found to be saturated or fell outside the field-of-view of the ACS $F606W$ image.  This effectively removed all of the bright foreground stars.  Visual inspection of objects 25 and 29 in Region 2 revealed that the $F814W$ detections were not exactly coincident with those in $F225W$, suggesting that two different objects -- one red and one blue -- had been observed in each case.  Whether the $F606W$ detections were of one, the other, or a blend of both the red and blue objects was unclear, due mostly to the transformation required to map the $F606W$ image to the WFC3 coordinate frame and the lower resolution of the ACS instrument.  The photometry of objects 25 and 29 has therefore not been plotted.  Object 24, also in Region 2, falls outside the ranges plotted in Fig.~\ref{fig:col_col_SSP} having colours inconsistent with SSPs and single star models (\S\ref{sec:single_star_fitting}) -- again suggesting the detection of different stars in different filters.

The photometry of Region 1 is colour-coded depending on the location of each object.  Objects coincident with strong levels of line-emission in the north are plotted in green, while those in the south that are associated with little or no discernible line-emission are plotted in red.  It is clear that objects in the north of Region 1 are systematically redder in ($F225W-F814W$) than objects in the south.  Assuming the two populations are of similar age, with similar intrinsic colours, suggests that the northern population is more heavily extinguished -- to be expected given its association with significant amounts of (line-emitting) gas.  It should also be noted that the ($F606W-F814W$) colour of some objects in Region 1 may be significantly bluer than their actual values due to the partial correction of emission line flux as described in \S\ref{sec:psf_fitting}.

Several points are immediately obvious from Fig.~\ref{fig:col_col_SSP}: (i) the colours of the vast majority of the NUV-detected objects are consistent with stellar populations less than 10 Myr old; (ii) {\em all} of the objects in Regions 1 and 2 of the inner filament have ($F225W-F814W$) colours consistent with such young stellar populations; (iii) objects in the northern part of Region 1 are more heavily extinguished than in the south; (iv) assuming foreground extinction, the bluest objects in the rest of the WFC3/ACS field may be on average {\em slightly older} than those in Regions 1 and 2; (v) the reddest objects [($F225W-F814W$) $>$1.5] are potentially globular clusters of intermediate age ($\sim$1 Gyr).  This includes object 81 (also labelled GC1 in Fig.~\ref{fig:inner_filament}) and object 109, the visibly extended cluster candidates \footnote{GC2 was not detected in $F225W$ and is therefore not included in our photometry.}.

It is important at this point to recognize the limitations of SSPs in modeling real stellar populations.  SSPs can be used to accurately model massive clusters ($\sim10^6$M$_{\odot}$), but are less effective when considering clusters of lower mass.  Model SSPs assume a fully populated IMF which is only true in the limit of an infinite number (and hence mass) of stars.  Stochastic sampling of the IMF in low mass clusters, particularly with reference to the most massive constituent stars, can lead to significant discrepancies between model colours and observed photometry \citep[e.g.][]{2002AJ....124..158D,2004A&A...413..145C,2004A&A...426..399J,2005A&A...443...79B}.  \citet{2004A&A...413..145C} have calculated minimum cluster masses ($\mathcal{M}^{min}$) below which cluster colours are expected to show significant dispersion in comparison to model SSPs.  At $\mathcal{M}^{min}$ this dispersion is reported to be at least 0.35 mag in any given filter.  $\mathcal{M}^{min}$ is dependent on both the age and metallicity of the cluster, and also the photometric band in which it is observed, with shorter wavelength observations permitting lower values of $\mathcal{M}^{min}$ .  \citet{2004A&A...413..145C} calculate that $U$-band photometry of a 4-10 Myr cluster can reach $\mathcal{M}^{min}$ of 10,000 - 3,000 M$_{\odot}$, dropping as low as 300 M$_{\odot}$ for a 100 Myr cluster.  Assuming a continuation of the trend to shorter wavelengths, our $F225W$ (NUV) photometry should enable even lower $\mathcal{M}^{min}$, but here we have opted to assume the $U$-band derived values of \citet{2004A&A...413..145C} as conservative limits.

Assuming an object to be a SSP, we can make crude estimates of its mass by taking rough age and extinction values from Fig.~\ref{fig:col_col_SSP} and comparing the absolute photometry to model SSPs.  Using ages of 2 to 5 Myr and extinction of E(B-V) = 0.35 we estimated object 15 -- one of the brightest in Region 1 -- to be roughly 1,800 - 2,700 M$_{\odot}$.  This is around 2 to 3 times lower than our $\mathcal{M}^{min}$ limit and we can therefore conclude that, for most objects, age and mass estimates derived from SSPs will have large uncertainties\footnote{In contrast, we derive a mass of 2.5$\times10^6$ M$_{\odot}$ for the large globular cluster candidate object 81 (also labelled GC1 in Fig.~\ref{fig:inner_filament}), assuming an age of 1 Gyr and foreground extinction of E(B-V) = 0.115}.

In an attempt to better utilize SSPs to estimate ages in Regions 1 and 2, we combined the individual object photometry within each region and compared the resultant colours to the entire grid of model SSPs.  By combining the photometry within each region we effectively increased the mass of the fitted stellar population.  However, we also implicitly assumed that all the objects within a region were coeval.  This assumption might be reasonable if star-formation was triggered by an event which traversed each region on a short timescale; e.g. a shock.  The estimated mass for Region 1 -- using ages of 2-5 Myr and extinction of E$(B-V)$ = 0.115 to 0.35 -- was found to be roughly 2,000 to 11,000 M$_{\odot}$, the range in mass due mostly to that in extinction.  The estimated mass for Region 2 was much lower at around 500 to 1,000 M$_{\odot}$, assuming only foreground extinction \citep[E$(B-V)$ = 0.115;][]{1998ApJ...500..525S}.  In either case the combined mass estimates still fall close to, or below the values of $\mathcal{M}^{min}$ discussed earlier, and we must therefore expect large uncertainties in our comparison with model SSPs.  To at least partially account for this in our SSP model fitting we included an uncertainty of $\pm$0.35 mag in each photometric band -- equivalent to the dispersion noted by \citet{2004A&A...413..145C} for photometry of clusters with masses of $\mathcal{M}^{min}$.  

The observed colours were compared to the entire grid of model SSP photometry, covering the range of ages and metallicities described at the beginning of this section.  Extinction was allowed to vary between 0.1 $\leq$ E$(B-V) \leq$ 1.0, with appropriate extinction coefficients derived for each filter assuming the reddening laws of \citet{1989ApJ...345..245C} and a model spectrum of B0 supergiant from \citet{2004astro.ph..5087C}.  The likelihood of each model ($\propto {\rm exp} -(\chi^2/2)$) was determined from the value of $\chi^2$, computed in the standard way.   Each parameter (age, extinction and metallicity) was then marginalised from the joint probability distribution to extract its one-dimensional probability density function (PDF).  This is the same method as employed by, for example, \citet{2007MNRAS.381L..74K}.

\begin{figure*}
\centering
\includegraphics[width=160mm]{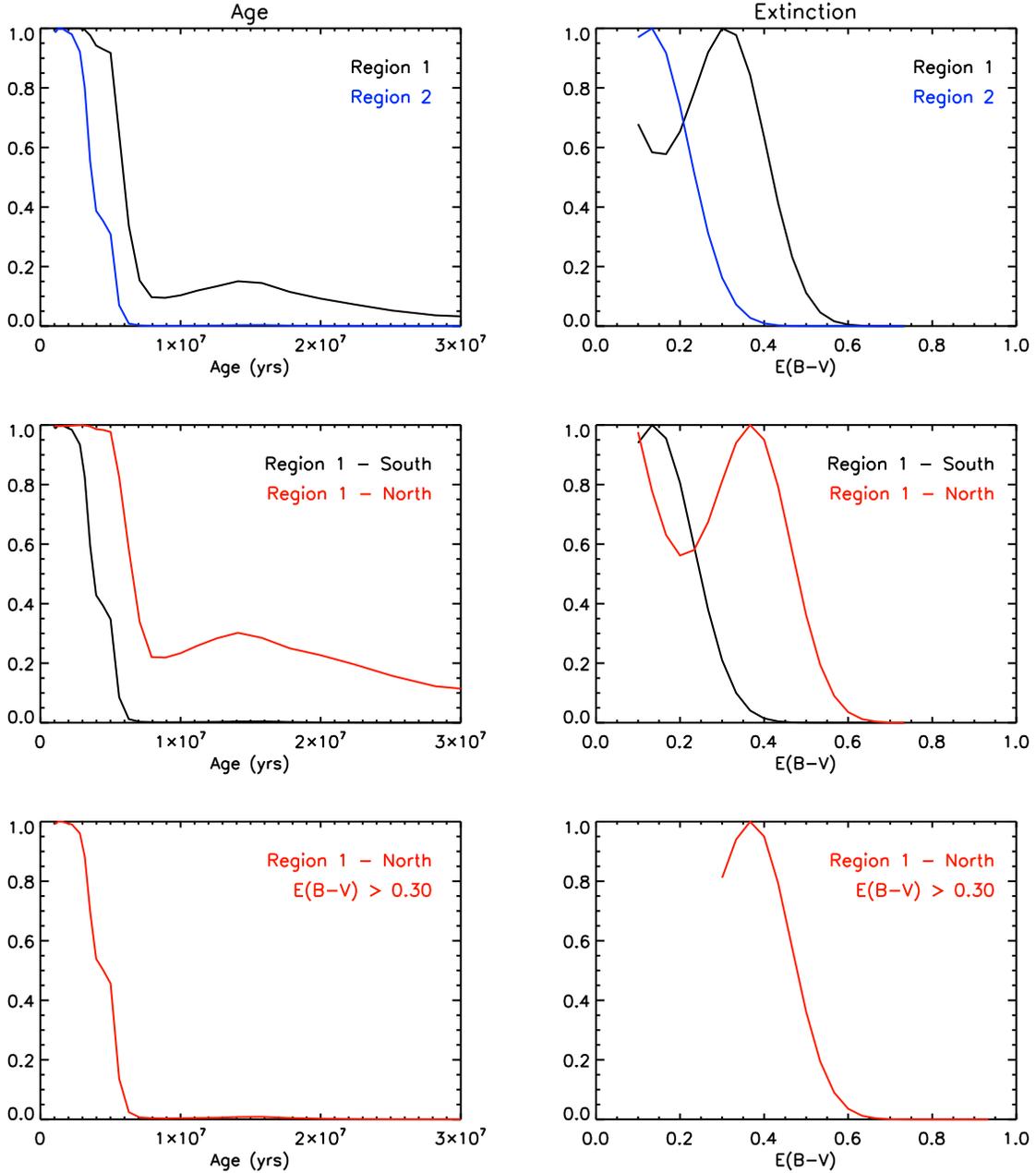}
\caption{Marginalised age (left) and extinction (right) PDFs for Regions 1 and 2 in the SW tip of the Cen A inner filament (see Fig.~\ref{fig:colour_regions}).  Observed HST photometry was fitted to colours of model SSPs, covering a range of ages, extinctions and metallicities (Age = $10^6$ to $10^9$ yrs; E(B-V) = 0.1 to 1.0; Z = 0.1 to 1.5 Z$_{\odot}$).  ({\bf Top panels:})  Both Regions 1 and 2 are best-fit by young stellar populations of only a few Myrs old.  However, the median age and extinction associated with Region 1 both appear to be higher than for Region 2.  ({\bf Middle panels:}) Separating Region 1 into Northern and Southern components, it appears that the Southern component actually has very similar age and extinction characteristics to Region 2, while the Northern component, which is associated with a bright HII region, most probably suffers higher levels of extinction.  ({\bf Bottom panels:})  Limiting the fitting of the Region 1 Northern component to models with E(B-V) $>$ 0.30, we find best-fit ages comparable to those of the Southern component and Region 2.  The marginalised metallicity PDFs (not shown) are flat, owing to the fact that the model colours for such young stellar clusters are almost completely independent of metallicity.}
\label{fig:PDFs}
\end{figure*}

The marginalised PDFs for Regions 1 and 2 are shown in the top panels of Fig.~\ref{fig:PDFs}.  The age-PDFs indicate that the stellar populations in both regions are very young (only a few Myrs old).  However, the main peak of the Region 1 age-PDF is significantly broader than that of Region 2.  The Region 1 age-PDF also possesses a lower-likelihood tail which extends to much older ages -- out to roughly 100 Myr.  Both of these features are most likely due to higher levels of extinction in Region 1, in particular in the northern part of this region, which is associated with significant line-emission.  This becomes clear when we split Region 1 into its northern and southern components, and fit each component separately (see middle panels of Fig.~\ref{fig:PDFs}) .  The southern component --- which is free from line-emission --- has similar age and extinction-PDFs to Region 2.  The northern component --- where we see intense line emission --- has a broader age PDF, and a second peak in the extinction PDF at higher reddening (E$(B-V) \approx$ 0.375 mag).  Higher extinction is exactly what one would expect in an HII region, owing to the presence of gas, and presumably dust.  Note that the higher extinction fits are associated with models of younger age, which are consistent with the best-fit ages from Region 2 and the southern component of Region 1.  The older ages are due to fits that involve lower assumed extinctions, such that the observed photometry appears intrinsically redder and hence older.  With more detailed spectral coverage we might have been able to break this {\em age-extinction} degeneracy using model fitting alone.  However, given its apparent association with an HII region, we can reasonably assume that higher extinction values are more likely in the northern part of Region 1.

From SSP fitting, we therefore conclude that the young stars in Regions 1 and 2 are most likely 1--4 Myrs old, and not older than $\sim$7 Myrs.

\subsubsection{Single stars}
\label{sec:single_star_fitting}

The individual source photometry was fitted to a grid of synthetic isochrones, again derived from the Padova stellar models \citep{2000A&AS..141..371G,2008A&A...482..883M}.  Each isochrone consisted of a set of single star models of the same age, but with different masses, and hence different luminosities and colours.  The model photometry grid included isochrones with ages from 10$^6$ to 10$^8$ yrs ($\Delta$ log(Age[yr]) = 0.1), and metallicities of  Z = 0.002 to 0.030 ($\Delta$ Z = 0.002).  

\begin{figure*}
\centering
\includegraphics[width=50mm,angle=90]{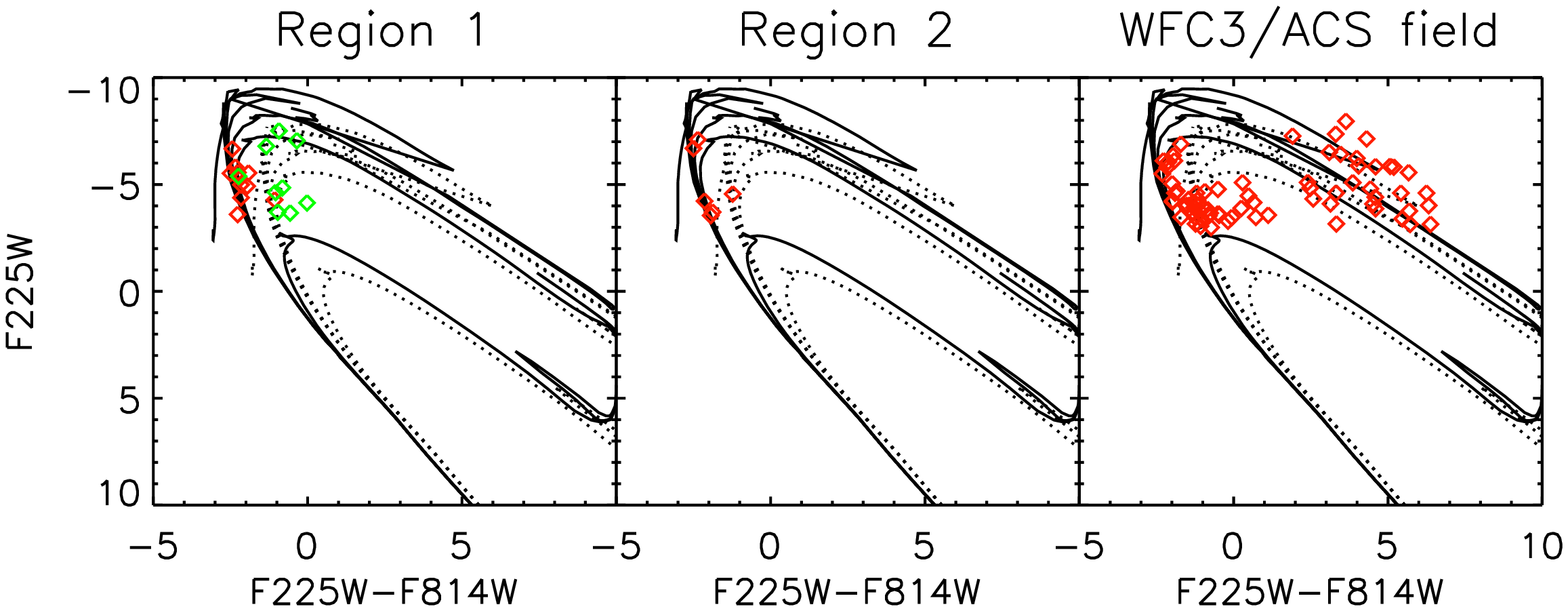}
\includegraphics[width=50mm,angle=90]{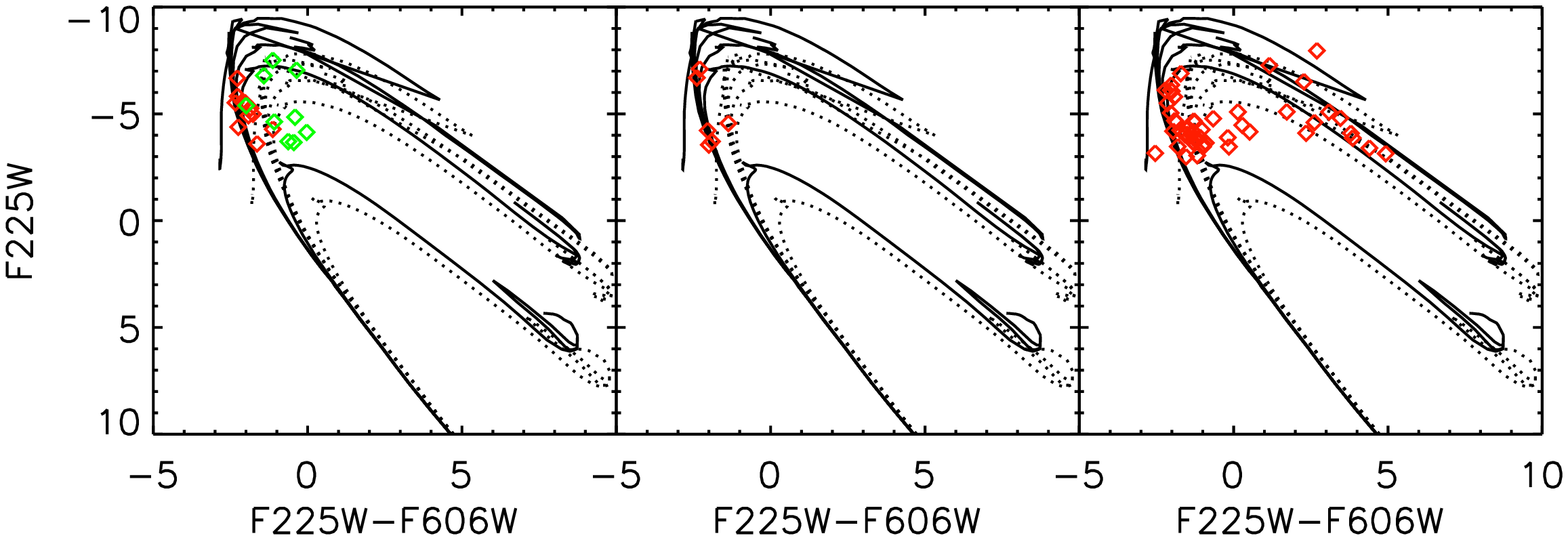}
\includegraphics[width=50mm,angle=90]{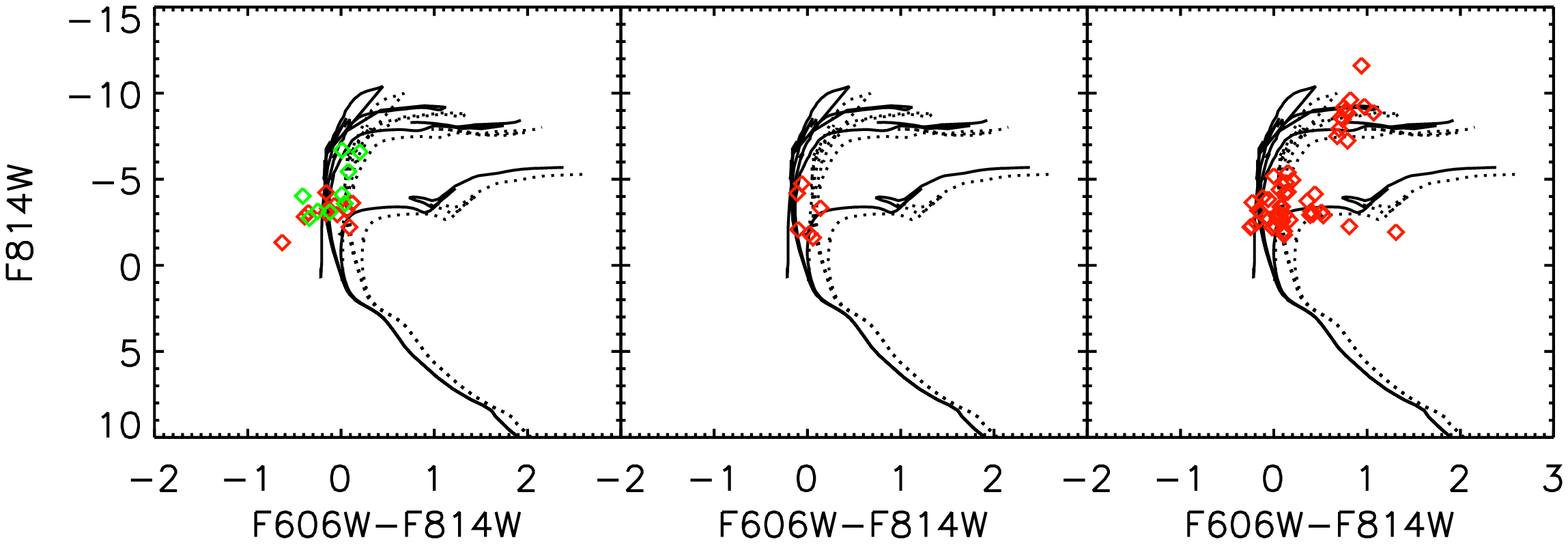}
\includegraphics[width=50mm,angle=90]{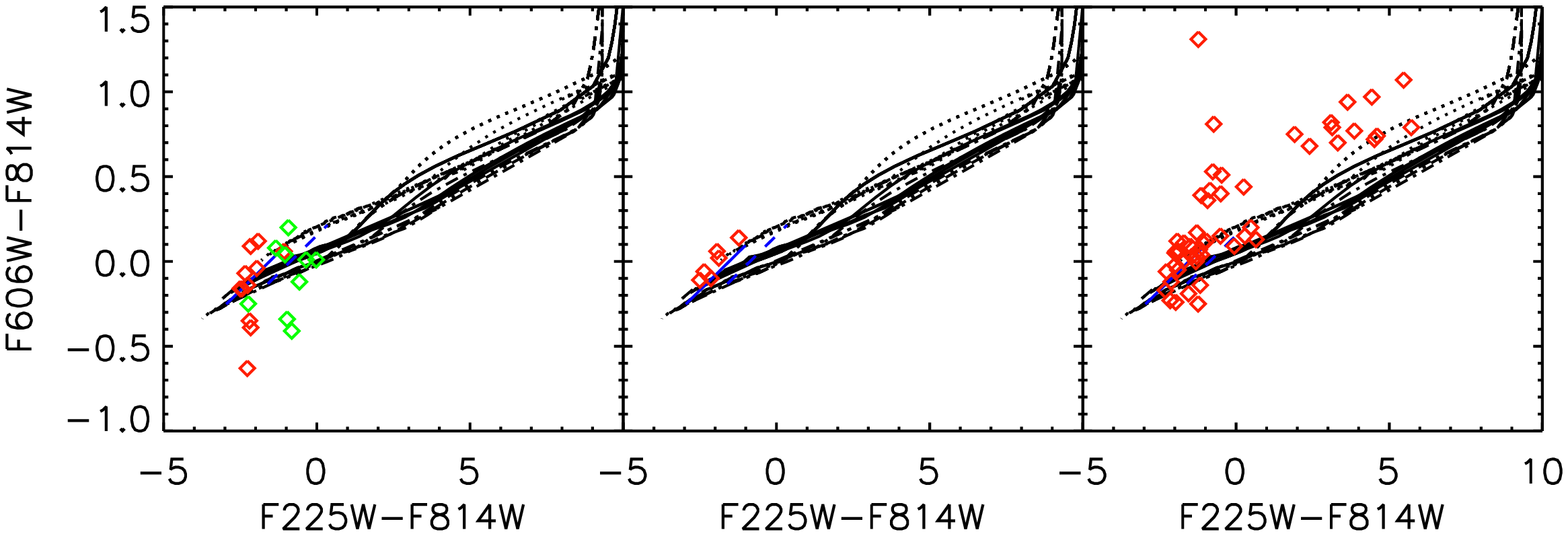}
\caption{Photometry of detected NUV-bright objects (red diamonds) overlaid on stellar isochrones: log(Age[yr]) = 6.0 to 7.0, in increments of 0.2 dex, and log(Age[yr]) = 8.0.  All the isochrones shown are of solar metallicity.  Colour vs magnitude plots show isochrones with two levels of extinction: foreground (solid line) and 3x foreground (dotted line).  Colour vs colour plot shows three levels of extinction: zero (dot-dashed line), foreground (solid line) and 3x foreground (dotted line). The photometry of objects in Region 1 has been split into two groups: northern objects associated with strong line-emission (green diamonds); southern objects associated with little or no line-emission (red diamonds).  The northern objects appear significantly redder than those in the south, most probably due to higher extinction. See text for detailed discussion.}
\label{fig:col_mag_isochrones}
\end{figure*}

Figure~\ref{fig:col_mag_isochrones} shows the observed HST photometry plotted in colour-vs-magnitude and colour-vs-colour space along with synthetic photometry from model isochrones.  The HST photometry is again divided into three groups, with Region 1 further divided into its northern component (green diamonds) -- which is coincident with significant line emission -- and its southern component (red diamonds).  It is useful to consider Fig.~\ref{fig:col_mag_isochrones} alongside Fig.~\ref{fig:NUV_object_posns}, which shows the position of each object in the WFC3/ACS field colour-coded relative to its ($F225W - F814W$) colour.  Note that not all the photometry listed in Table~\ref{tab:phot_table} is plotted in Fig.~\ref{fig:col_mag_isochrones}, as some objects did not have a detection, or were saturated in one or both of the $F606W$ and $F814W$ images.  The plotted isochrones are of roughly solar metallcity (Z = 0.020) with log(Age[yr]) = 6.0 to 7.0, in increments of 0.2 dex, and log(Age[yr]) = 8.0.  Two levels of extinction are illustrated in the colour-vs-magnitude plots -- E$(B-V)$ = 0.115 and 0.345 -- while three levels are shown in the colour-vs colour plots -- E$(B-V)$ = 0.0,  0.115 and 0.345.

We again see that objects in the northern component of Region 1 are systematically redder than those in the south.  This reddening is especially pronounced in the NUV-optical colours, and is most easily explained as being due to excess extinction associated with the line-emitting gas in the north of Region 1.  The objects in Regions 1 and 2 are most consistent with single star models of ages $<$ 10 Myr.  The majority of blue objects in the rest of the field appear to be relatively consistent with similarly young stars, although there is some indication that the absolute $F225W$ magnitudes are on average fainter than those in Regions 1 and 2 perhaps suggesting slightly older ages. 

Many of the reddest obects --- ($F225W - F814W$) $>$ 1.5 --- are saturated in $F606W$ and are most likely foreground stars.  By assuming distance moduli consistent with Milky Way stars along the line-of-sight, all of these objects can be made to lie along the main sequence (roughly 14 magnitudes fainter than currently plotted object photometry).  Other red objects, which have good, non-saturated detections in all filters, are found to have colours inconsistent with single stars (see the colour-vs-colour plot in Fig.~\ref{fig:col_mag_isochrones}).  These same sources were discussed in the previous section on SSP fitting where we suggested they are GC candidates of $\sim$ 1 Gyr old.  The distribution of the red objects across the field-of-view lends further support to our conclusions.  If these objects were a population of red supergiants in Cen A we would expect them to be associated with regions of very recent star-formation -- i.e. the youngest stars in the field.  Since they are in fact spread quite evenly across the field, it is more likely that they are part of the old stellar background in Cen A, or foreground stars in our own Galaxy.

Two colours --- ($F225W - F606W$) and ($F606W - F814W$) --- as well as the absolute magnitude, M$_{F814W}$, of each of the NUV-bright sources in Regions 1 and 2 were fitted to the entire grid of model isochrones, while extinction was allowed to vary between 0.1 $\leq$ E$(B-V) \leq$ 1.0 mag.  Note again that foreground extinction towards Cen A is measured to be E$(B-V)$ = 0.115 \citep{1998ApJ...500..525S}.   The likelihood of each model fit ($\propto {\rm exp} -(\chi^2/2)$) was again determined from the value of $\chi^2$.  By fitting M$_{F814W}$ we were able to break some of the degeneracy in colour-colour space --- i.e. distinguish stars of the same colour but different luminosity (and hence age).  Note that the absolute magnitude uncertainty included a contribution from the distance modulus of Cen A of $\pm0.24$ mag.

\begin{figure*}
\centering
\begin{tabular}{cc}
\vspace{2mm}
\includegraphics[width=65mm]{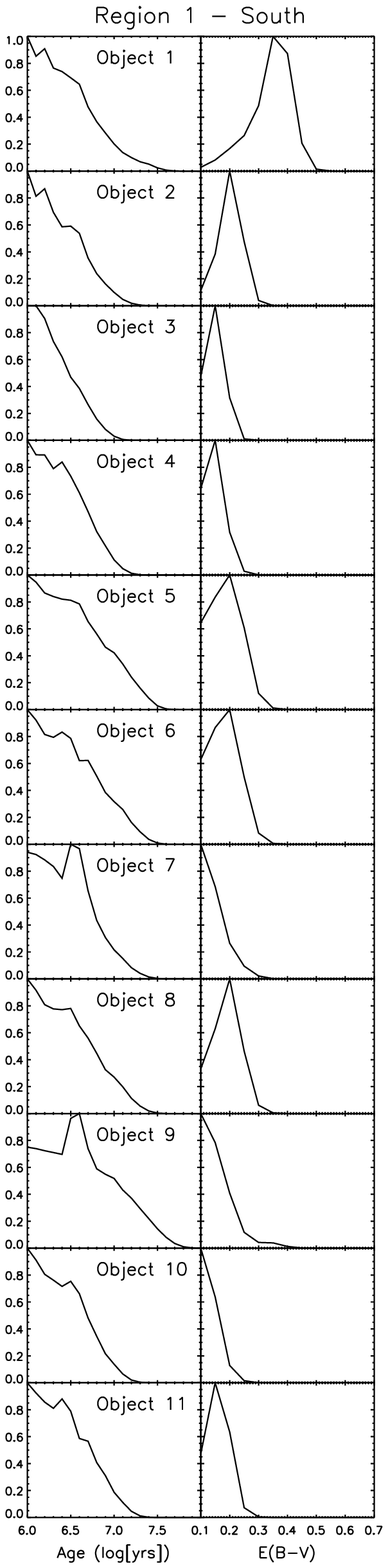}&
\includegraphics[width=65mm]{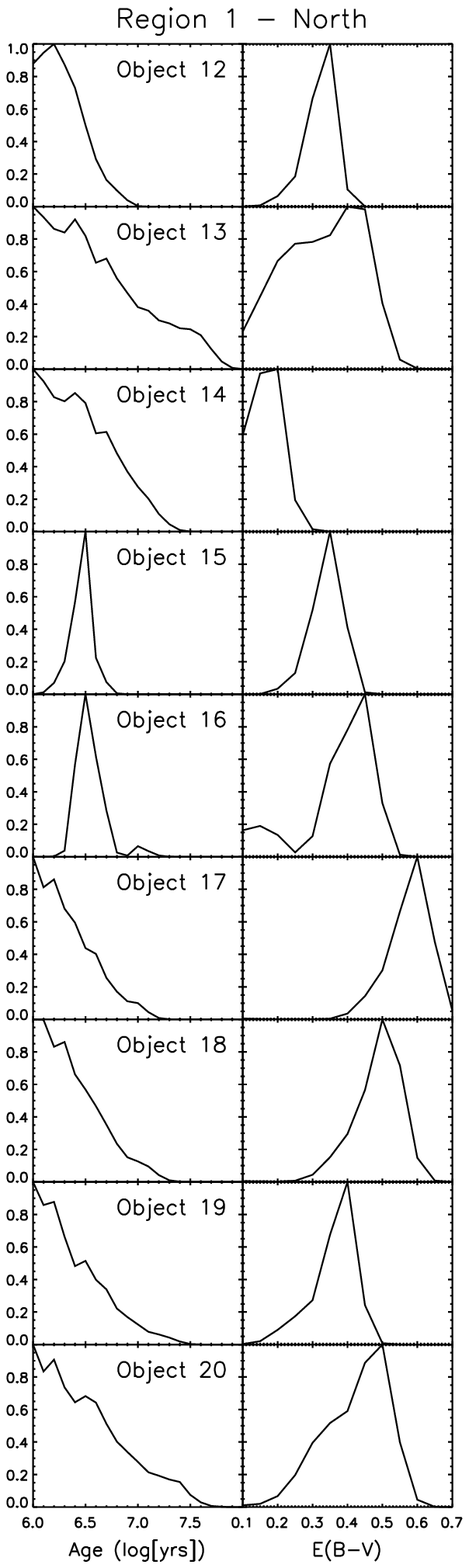}\\
\end{tabular}
\caption{Marginalized age and extinction PDFs for NUV bright objects in Region 1 of the inner filament south-west tip.  See Section~\ref{sec:single_star_fitting} for details.  Object ID's correspond to those shown in the top-right panel of Fig.~\ref{fig:colour_regions}.}
\label{fig:region1_pdfs}
\end{figure*}

\begin{figure}
\centering
\includegraphics[width=65mm]{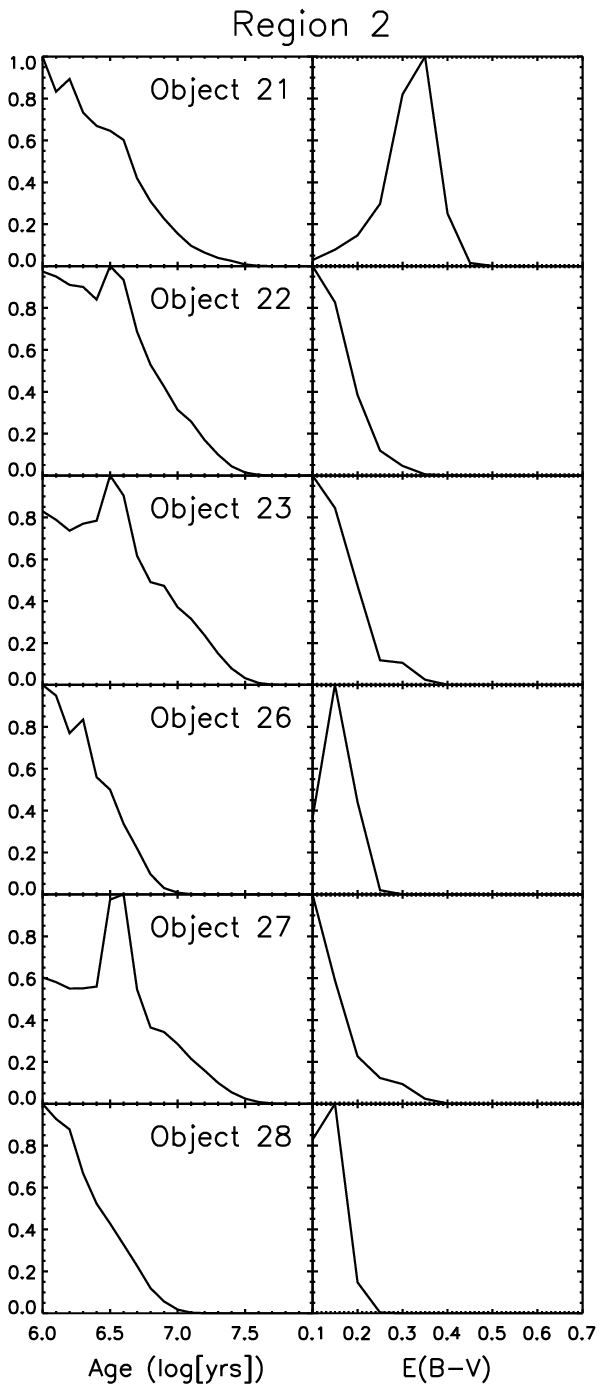}
\caption{Marginalized age and extinction PDFs for NUV bright objects in Region 2 of the inner filament south-west tip.  See Section~\ref{sec:single_star_fitting} for details.  Object ID's correspond to those shown in the bottom-right panel of Fig.~\ref{fig:colour_regions}.}
\label{fig:region2_pdfs}
\end{figure}

Figs.~\ref{fig:region1_pdfs}\,\&\,\ref{fig:region2_pdfs} show marginalized age and extinction-PDFs for all of the objects in Regions 1 and 2 (see Fig.~\ref{fig:colour_regions}).  In Fig.~\ref{fig:region1_pdfs} the objects of Region 1 are separated into two groups corresponding to the northern and southern components discussed in Section~\ref{sec:SSP}.  Note that the objects in the north are generally associated with strong \Ha\/ emission, while those in the south are not, as can be seen in Fig.~\ref{fig:colour_regions}.  Our single star model fits indicate systematically higher values of extinction in the northern component of Region 1 when compared to the south, which is consistent with the result from SSP fitting in the previous section.  The majority of objects in Region 2 are best-fit with low values of extinction, again consistent with the SSP fitting results.

The marginalized age PDFs suggest that all of the NUV bright sources in Regions 1 and 2 are $\lesssim$10-20 Myrs old, assuming that they are single stars.  This upper age limit is in line with those derived from SSP fitting, although somewhat less constraining.  Overall, the single star fits appear to corroborate the age and extinction results derived from SSP fitting.

\section{Triggered star-formation?}
\label{sec:interpretation}

The multi-wavelength picture for the inner \Ha\/ filament is an interesting one. Starting at the NUV-bright south-west tip (corresponding to a location near Region A of \citealt{1991MNRAS.249...91M}) and moving north-east, the UV emission quickly disappears. Instead, diffuse, gradually decreasing X-ray emission is observed to be associated with Regions B and C \citep{2004ApJ...617..209E}, and no X-rays at all are found further along the filament. The broad-band optical emission paints a similar picture to that in the near UV, with discernible stellar clusters only found coincident with the strong NUV-flux. This strongly suggests that star-formation (recent or otherwise) has only occured at the south-west tip of the filament, the region closest to the radio lobes.

\citet{1993ApJ...414..510S} suggested that the observed \Ha\/ emission can arise as a result of interaction of dense molecular clouds with AGN radio plasma. In this picture, the radio cocoon induces supersonic turbulence in the gas, ablating it from the dense cloud. Supersonic clumps produce strong extreme UV and soft X-ray photons as they collide, giving rise to the high-excitation lines. Low-excitation lines come from shocks in the dense gas. Such multi-phase gas is a natural consequence of both the Kelvin-Helmholtz instability, and ablation of the outer parts of gas clouds.

The \citeauthor{1993ApJ...414..510S} model successfully reproduces both the high and low excitation lines, as well as the H$\beta$ luminosity observed by \citet{1991MNRAS.249...91M}. Moreover, the predicted X-ray luminosity in the filament is also consistent with the observed value \citep{2004ApJ...617..209E}. However, there are two major problems with the model. Firstly, the observed X-ray emission is much more extended than models predict. Secondly (and this is the fundamental issue), \citeauthor{1993ApJ...414..510S} predicted ``the presence of a radio jet in the vicinity of the inner filaments''. As seen in Figure~\ref{fig:CenA_large}, no such jet is observed.

Our WFC3 NUV observations shed new light on this problem. In this section, we propose an alternative scenario for formation of the filaments, namely shock heating and ablation of a gas cloud by a weak bow shock associated with the radio lobes.

\subsection{Ablation of gas clumps}
\label{sec:dragging}

\begin{figure*}
\centering
\begin{tabular}{cc}
\includegraphics[width=80mm]{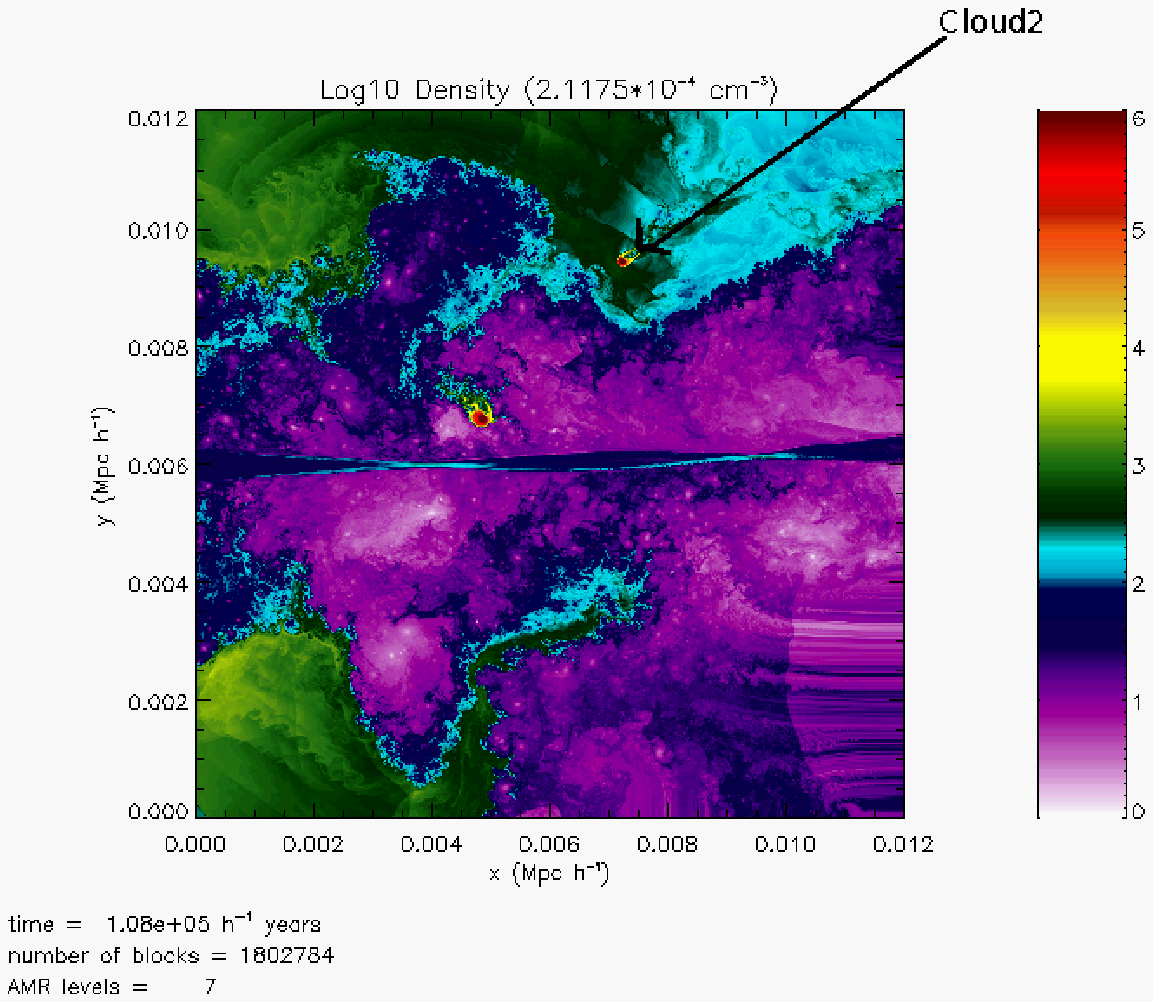}&
\includegraphics[width=80mm]{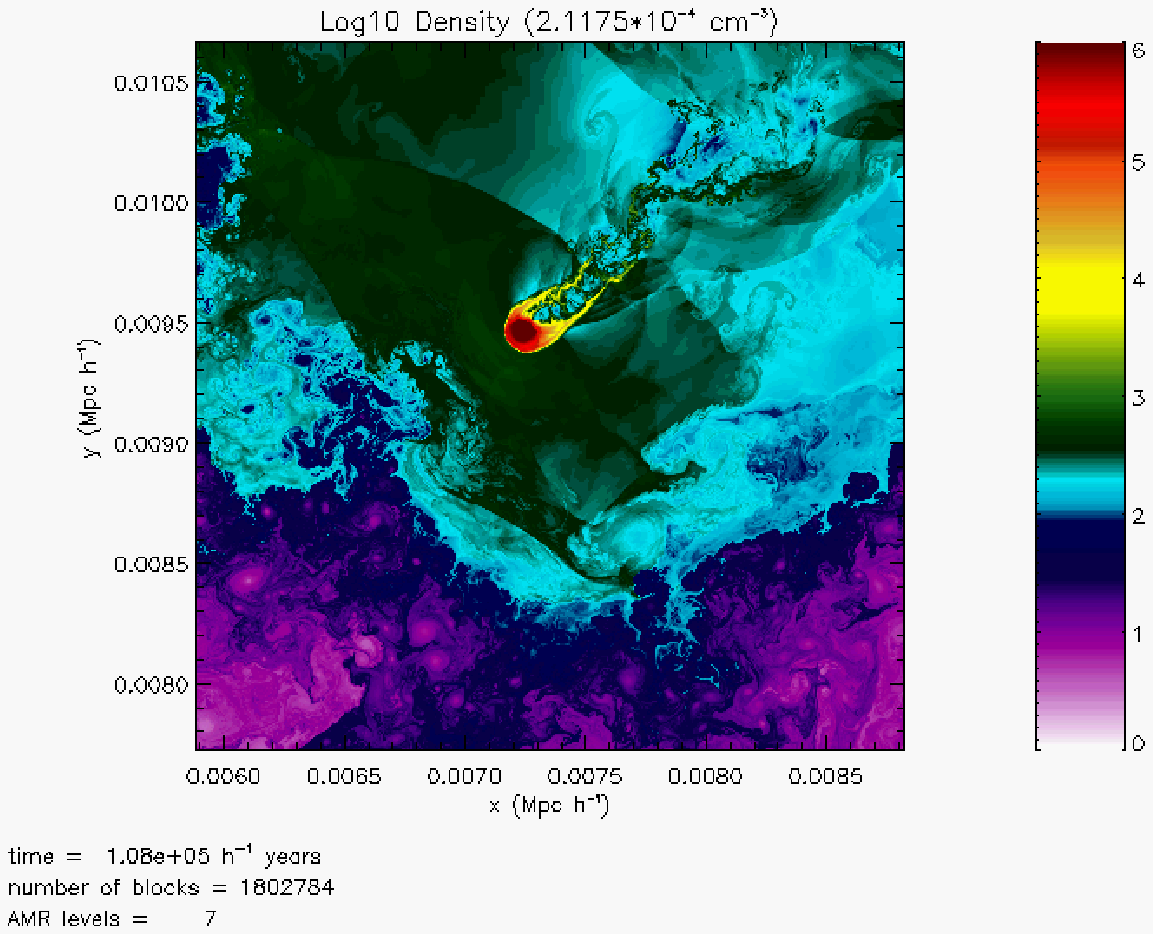}
\end{tabular}
\caption{Density maps from a two dimensional hydrodynamical simulation showing the Cen A AGN-inflated radio cocoon expanding supersonically into intragalactic gas. {\bf (Left panel:)} A mildly supersonic bow shock is induced ahead of the expanding radio cocoon, which overruns a molecular gas cloud, denoted ’Cloud2’. The underdense (with respect to the intragalactic medium) radio cocoon is shown in purple. Ambient density of the undisturbed gas corresponds to green colours. The expanding bow shock is outside the region shown. {\bf (Right panel:)} The passage of the shock triggers star-formation in the densest part of the cloud (red and yellow colours), and ablates the more diffuse gas, dragging it out along the direction of shock propagation. The ablated material is shock ionised, giving rise to the emission-line ﬁlament. A detailed 3D simulation of the formation of the Cen A inner filament will be presented in an upcoming paper (Antonuccio-Delogu et al. in prep.).}
\label{fig:simulation}
\end{figure*}

The basic picture we propose is illustrated in Fig.~\ref{fig:simulation} and is described as follows. An AGN-inflated radio cocoon expands supersonically into intragalactic gas. The resultant mildly supersonic bow shock overruns a molecular gas cloud. The densest parts of the cloud are radiative, and the passage of the shock triggers the star-formation observed at the south-west tip of the inner filament. The more diffuse parts of the cloud are ablated, and compressed as they are dragged upwards by the shock front. The morphology of the resulting wake strongly depends on local microphysics (eddies and Kelvin-Helmholtz instabilities), and evolves with time. Shocks similar to those proposed by \citet{1993ApJ...414..510S} give rise to the \Ha\/ emission further up the filament. The shock-heated gas will also emit in the X-rays, giving rise to spatially distributed emission.

\subsubsection{Filaments}

Filamentary structures formed by uplifting of gas clumps have been observed in other environments. For example, \citet{2003MNRAS.344L..43F} report observations of \Ha\/ filaments in the Perseus cluster, and around NGC\,1275 in particular. These authors argue that such structures are a natural consequence of the buoyant rise of an underdense bubble of radio plasma through the cluster gas. Essentially, these filaments correspond to gas ``falling off'' the bubble as it rises. It seems plausible that a weak shock could play a similar role in Centaurus~A.

Another possibility is that gas clumps were already lined up along the filament before something triggered the \Ha\/ emission.  This can happen as a result of, for example, a thermal instability \citep{2010ApJ...720..652S}. Our observations argue strongly against this possibility. The warm interstellar medium is multi-phase \citep[e.g.][]{2007ApJS..173...37S}, and any shocks required to give rise to \Ha\/ emission would necessarily also give rise to star-formation in the densest parts of the ISM. The resolution and sensitivity of WFC3 is sufficiently high to observe compact groups of (or even single) massive stars, and thus there is no surface brightness limit to consider. In other words, any young ($\leq 10$~Myrs) massive stars--- even if formed in very small numbers--- should be observed in the near UV. Similarly, any older stars should be picked up in the $i$-band images. The fact that we don't see any stellar emission anywhere--- apart from the south-west tip of the filament--- suggests that no dense gas clumps have been overrun in those regions. In other words, our WFC3 observations strongly favour an external origin for the emitting gas. It is worth noting that \citet{2003MNRAS.344L..43F} also conclude that it is very unlikely that the cold gas clumps giving rise to the \Ha\/ filaments in the Perseus cluster are formed {\it in situ} by cooling out of the hot gas.

\subsubsection{\Ha\/ in the star forming region}

If the above picture is correct, the lack of shock-heated X-ray gas at the south-west tip, 
together with the young stellar population, suggests that only the densest gas remained unablated. Consequently, much of the observed \Ha\/ emission must be due to the photoionizing UV-flux associated with star-formation, in contrast to the rest of the filament.

\citet{1998ApJ...509...93W} relate \Ha\/ surface brightness to the density of photoionized electrons $n_e$. Rearranging their Equation~2, we get
\begin{equation}
	n_e = \left( \frac{\Sigma_{\rm H\alpha}}{5.9 \times 10^{-14} {\rm\,ergs\,s^{-1}\,cm^{-2}\,arcsec^{-2}}} \right) U^{-1} e_{\rm H\alpha} {\rm\, cm^{-3}}
\label{eqn:n_e}
\end{equation}
where the ionization parameter $U=4.22 \times 10^{-4} r^{-1.72}$ is empirically found from the ratio $r = \frac{\rm [S\,II] \lambda \lambda 6716,6731}{{\rm H} \alpha}$, and $e_{\rm H\alpha}$ is the ratio of intrinsic to observed surface brightness (i.e. an extinction correction in linear flux).

 \citet{1991MNRAS.249...91M} give an average \Ha\/ surface brightness value for the entirety of Region A (some 30 arcsec$^2$ in area) of 1.08 $\times 10^{-15}$ ergs\,s$^{-1}$\,cm$^{-2}$\,arcsec$^{-2}$.  We find a similar value (1.17$\times 10^{-15}$ ergs\,s$^{-1}$\,cm$^{-2}$\,arcsec$^{-2}$) by performing photometry of the entire region in the continuum-subtracted WFC3 $F657N$ image, using an 80 pixel (3.2$\arcsec$) radius aperture.  However, around 40 percent of the total \Ha\/ emission in Region A is concentrated into an area of just 0.5 arcsec$^2$ (0.4$\arcsec$ radius) centred on the main star-cluster visible in Figs.~\ref{fig:colour_ABC}\,\&\,\ref{fig:colour_regions}.  The \Ha\/ surface brightness in this small area is obviously much higher than the average value quoted by \citet{1991MNRAS.249...91M}

We used a 10 pixel (0.4$\arcsec$) radius aperture to measure the \Ha\/-flux coincident with the main star-cluster in the continuum-subtracted WFC3 $F657N$ image.  Correcting for the instrument and filter throughputs, we calculated a mean surface brightness within the aperture of 2.90$\times 10^{-14}$ ergs\,s$^{-1}$\,cm$^{-2}$\,arcsec$^{-2}$, which is almost 30 times higher than the average value for Region A.  Note that neither this, nor the \citeauthor{1991MNRAS.249...91M} value as quoted, have been corrected for extinction.

Literature values--- and our best-fit SSP models (Section~\ref{sec:sf})--- suggest modest extinction values between E(B-V) = 0.1 and 0.3 mag, yielding $e_{\rm H\alpha} \approx$ 1.3 to 1.9. The line ratios listed in Table~3 of \citet{1991MNRAS.249...91M} allow us to calculate $r=0.22$ for Region A, and thus $U=5.71 \times 10^{-3}$. From Eqn.~\ref{eqn:n_e} we then find that ionized gas densities of $n_e \approx$ 112 to 164~cm$^{-3}$ (depending on the value of  $e_{\rm H\alpha}$) are required to give rise to the observed \Ha\/ surface brightness, as measured in our 10 pixel aperture.  The ionization parameter $U$ is related to the total ionizing photon flux via $\Phi = U n_e c / 4$ \citep{1998ApJ...509...93W}. An ionizing flux of greater than $(4.8-7.0) \times 10^{13}$~photons\,s$^{-1}$\,m$^{-2}$ is therefore required. 

We can estimate the ionizing photon flux of our observed cluster from model SSP spectra.  In Section ~\ref{sec:SSP} we found the most likely age of the recent star-formation to be 1--3 Myrs.  Assuming that a value of 2 Myrs is appropriate, the mass of the cluster can be estimated by scaling a model 2 Myr old SSP to match the observed $F814W$ (10 pixel aperture) photometry.  This yields a cluster mass of roughly 5200~\Msun\/.

By integrating the model spectrum of a 2 Myr old SSP \citep[Starburst99; see][]{1999ApJS..123....3L} shortward of the 911~\AA~threshold for hydrogen ionization, and subsequently scaling to the cluster mass and surface area of the emitting region (i.e. a sphere of the same physical radius as our photometry aperture; 0.4$\arcsec$ = $2.20\times10^{17}$\,m), we predict $\Phi = 2.75\times10^{14}$~photons\,s$^{-1}$\,m$^{-2}$.  This is in excess of the value required to explain the observed \Ha\/ emission, suggesting that it can be explained purely by photoionization by young stars, in this region at least. 

\begin{figure*}
\centering
\includegraphics[width=100mm]{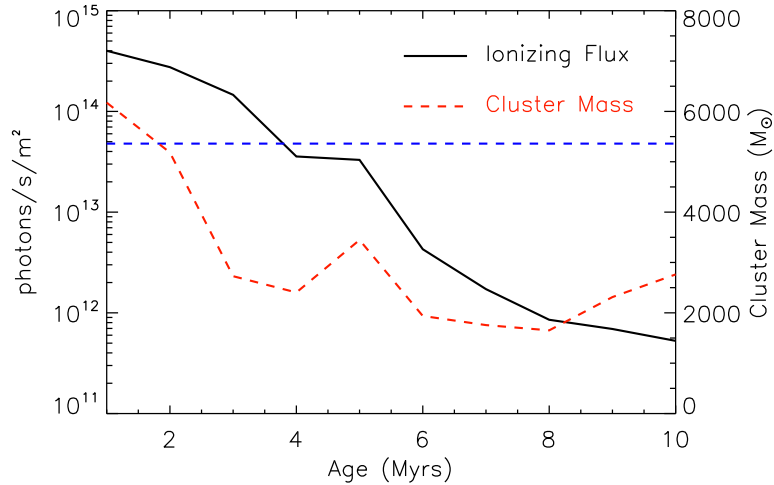}
\caption{Ionizing photon fluxes and cluster mass estimates vs age of the main star cluster in Region 1.  This cluster is embedded in an HII region in the Northern component of Region 1 (see Fig.~\ref{fig:colour_regions}).  Masses are estimated by scaling model SSPs of different ages to match $F814W$ photometry of the observed cluster.  Ionizing photon fluxes are calculated by integrating model SSP spectra shortward of 911~\AA~(the threshold for hydrogen ionization), which are scaled to match the estimated cluster masses.  The observed \Ha\/ surface brightness in the immediate vicinity of the cluster requires an ionizing photon flux of at least $4.79 \times 10^{13}$~photons\,m$^{-2}$\,s$^{-1}$ (blue dashed line) suggesting the stellar population must be younger than 4 Myrs.}
\label{fig:UVfluxPopulation}
\end{figure*}

This approach can be used to estimate the ionizing photon flux for a range of assumed cluster ages, and therefore provides an alternative method of estimating the age of the stellar population from the \Ha\/ surface brightness alone.  As Figure~\ref{fig:UVfluxPopulation} shows, stellar ages in excess of $4$~Myrs cannot produce the required ionizing flux.  Note that the mass of the cluster--- calculated from its $F814W$ photometry--- varies with assumed age.

We note that the above calculation only presents a plausibility argument for a significant contribution from newly-formed stars to the ionizing flux. Shocks will almost certainly play a role, particularly at greater radial distance from the young stars, and the relative contributions of the two components can only be decoupled with detailed high-resolution spectroscopy. In Section~\ref{sec:soundWaves} we show that this assumption leads to a prediction for the diffuse gas densities that are in good agreement with an independent method.

\subsection{Weak shock}
\label{sec:soundWaves}

The lack of a radio jet in the vicinity of the inner filament suggests that the origin of \Ha\/ emission is quite different from that in the outer filament. As outlined above (Section~\ref{sec:dragging} and Fig.~\ref{fig:simulation}), a weak shock that shocks and ablates a gas cloud located at the south-west tip of the inner filament could account for the observed features. Such a shock could arise from an overpressured cocoon of radio plasma expanding into the ISM. It is intriguing to note that an H\,I clump is detected in both absorption and emission \citep{2010A&A...515A..67S} at the north-eastern edge of the inner northern radio lobe--- the part of the radio emission closest to our filament.  This is also the region where the radio jet appears to bend. The very location of the H\,I cloud suggests that the jet could be deflected by the cloud \citep{2010NewA...15...96G}, as often seen in simulations \citep[e.g.][]{2007ApJS..173...37S,2011MNRAS.411..155G}. The resultant reorientation of the hotspot (where the jet terminates) would change the dynamics of cocoon backflow, and allow the bow shock far away from the hotspot to dissipate.  A detailed 3D simulation of the formation of the Cen A inner filament will be presented in a subsequent paper (Antonuccio-Delogu et al. in prep.).

\subsubsection{X-ray emission}
\label{sec:x-ray_emission}

{\it Hot gas density}

A shock needs gas to propagate through, and we therefore expect diffuse X-ray emission to be present between the radio lobes and the inner filament. \citet{2002ApJ...577..114K} find such diffuse emission with {\it Chandra}, however their maps are adaptively smoothed to 4.'2 resolution. \citet{2004ApJ...617..209E}, on the other hand, do not detect any diffuse X-rays in the region of interest. A lack of observed X-ray emission between the radio lobes and inner filament would argue against our interpretation, unless the gas is too diffuse to be detected in the X-rays. In the first instance, we follow the \citet{2003MNRAS.344L..48F} analysis of the Perseus cluster in assuming that the hot ISM is in pressure equilibrium with the \Ha\/ emitting gas near the south-west tip of the filament. For $10^7$~K X-ray emitting gas and $10^4$~K \Ha\/ gas this yields a density of $n_X \approx 10^{-3} n_e$.  While $n_e \approx$ 100 -- 150~cm$^{-3}$ immediately surrounding the main star-cluster in the south-west tip, the average for Region A is $n_e \approx 4$~cm$^{-3}$. The density of X-ray emitting gas is therefore estimated to be $n_X \sim 4 \times 10^{-3}$~cm$^{-3}$.

{\it Dynamical considerations}

Alternatively, average X-ray gas density can be estimated from radio source dynamics. Analytical models of \citet{1997MNRAS.286..215K} and \citet{1997MNRAS.292..723K} relate the size and synchrotron luminosity of double-lobed radio sources to physical parameters such as jet power, source age and the atmosphere into which the radio source is expanding. We follow an approach similar to that outlined by \citet{2008MNRAS.388..625S} to infer the density of the hot gas.

The age of inner lobes of Centaurus A is $t_{\rm age}=5.6-5.8$~Myrs, and their radio luminosity is $(1.05 \pm 0.16) \times 10^{24}$~\WHz\/ at 327~MHz \citep{2000A&A...355..863A}. Such young lobes are not expected to suffer significant synchrotron or Inverse Compton losses. In this case, the models of \citet{1997MNRAS.286..215K} and \citet{1997MNRAS.292..723K} give a closed-form expression for radio luminosity:
\begin{eqnarray}
  L_{\rm radio} & = & 8.0 \times 10^{27} \left( \frac{\nu}{\rm GHz} \right)^{-0.57} \left( \frac{R_T}{3(5+4R_T^2)-4-\beta} \right)^{1.19} \left( \frac{\rho_{\rm core}}{10^{-22}\,{\rm kg\,m}^{-3}} \right)^{0.595} \nonumber\\
  & & \times \left( \frac{Q_{\rm jet}}{10^{36}\,{\rm W}} \right)^{1.19} \left( \frac{D_{\rm source}}{\rm kpc} \right)^{0.62} \left( \frac{D_{\rm source}}{R_{\rm core}} \right)^{0.595 \beta}
\label{eqn:LradioLossless} 
\end{eqnarray}
Here, $D_{\rm source}$ is the observed separation of the two lobes, $\nu$ is the observing frequency, $R_T$ the axial ratio (length/width) of the radio cocoon, $Q_{\rm jet}$ is the jet power supplied to each lobe (equal to half the total jet power), and $\rho_{\rm core}$, $R_{\rm core}$ and $\beta$ describe the density profile of the hot gas, taken to follow a power law $\rho(r) = \rho_{\rm core} \left( \frac{r}{R_{\rm core}} \right)^{-\beta}$. All the other jet and environment parameters (power-law exponent for the electron energy spectrum at hotspots, adiabatic indices for the cocoon and magnetic field) are taken from Case~3 of \citet{1997MNRAS.292..723K}.

A second equation is required to close the relationship between $Q_{\rm jet}$ and the density profile. The dynamical model of \citeauthor{1997MNRAS.286..215K} relates the age and size of the radio source to jet power and environment, via:
\begin{equation}
  D_{\rm source} = 2 c_1 R_{\rm core}^{\beta/(5-\beta)} t^{3/(5-\beta)} \left( \frac{Q_{\rm jet}}{\rho_{\rm core}} \right)^{1/(5-\beta)}
\label{eqn:Dsource}
\end{equation}
where the constant $c_1 = \left[ \left( \frac{32}{27 \pi} \right) \left( \frac{R_T^4 (5-\beta)^3}{3(5+4R_T^2)-4-\beta} \right) \right]^{1/(5-\beta)}$ is of order unity.

The inner lobes of Centaurus~A have $D_{\rm source} \approx 10$~kpc, $R_T \approx 4.5$. Upper and lower limits on the gas density can then be obtained by assuming a very steep or very shallow gas density profile. Setting $\beta=0$ (a constant density profile), Eqns~\ref{eqn:LradioLossless} and \ref{eqn:Dsource} yield $\rho_{\rm core} = 4 \times 10^{-23}$~kg\,m$^{-3}$, corresponding to $n_X = 0.04$~cm$^{-3}$. Alternatively, setting $\beta=1.5$ and $R_{\rm core}=1$~kpc (approximately the transverse size of the H\,I gas cloud near the central engine) yields $\rho_{\rm core} = 1.9 \times 10^{-23}$ or $n_{\rm core} = 0.019$~cm$^{-3}$. At the lobes, this density drops by a factor $\left( D_{\rm source} / (2 R_{\rm core}) \right)^{-\beta}$, giving $n_X = 1.7 \times 10^{-3}$~cm$^{-3}$. These values are consistent with the equilibrium-derived value of $n_X = 4 \times 10^{-3}$~cm$^{-3}$. The implied jet powers of $(1 - 8) \times 10^{33}$~W are consistent with the lower limit of $8 \times 10^{32}$~W estimated from X-ray observations \citep{2009ApJ...698.2036K}.

We can also use the size and age of the radio lobes to estimate the expansion speed of the cocoon. Given the above values, the average cocoon expansion speed is $v_c \approx 510-730$ km\,s$^{-1}$, depending on the assumed density profile. Self-similar models \citep[e.g.][]{2002MNRAS.335..610A} predict the bow shock to expand a factor of 1.2 faster ($v_c \sim 600-900$~km\,s$^{-1}$). For comparison, X-ray temperatures of $\sim 0.7$~keV \citep[e.g.][]{2008ApJ...677L..97K} yield sound speeds of 430~km\,s$^{-1}$. Thus the picture of a mildly supersonic bow shock is a plausible one.

{\it Surface brightness}

Given a density and temperature, it is possible to derive the X-ray surface brightness expected from bremsstrahlung cooling. For a slab-like geometry, we have:
\begin{equation}
  \Sigma_X = 9.23 \times 10^{-18} \left( \frac{n_X}{{\rm cm}^{-3}} \right)^2 \Lambda(T_X) D \left( \frac{\theta_d}{\rm arcsec} \right) {\rm\,ergs\,s^{-1}\,cm^{-2}\,arcsec^{-2}}
\label{eqn:Sigma_X}
\end{equation}
where $D=3.68$~Mpc is the distance to Centaurus~A, the cooling term for $T_X=10^7$~K is $\Lambda \sim 2.5 \times 10^{-23}$~ergs\,s$^{-1}$\,cm$^3$ \citep{1993ApJS...88..253S}, and $\theta_d$ is the depth (in arcsec) of the X-ray emitting region along the line of sight. The size of the emitting region is $d \sim \left( \frac{3 M_{\rm gas}}{4 \pi \rho_{\rm core}} \right)$. The spheroid mass of Centaurus~A is $6.5 \times 10^{10}$~\Msun\, and the absolute upper limit on hot gas mass is 10\% of this value, yielding $d \leq 35$~kpc at the equilibrium density of $n_X = 4 \times 10^{-3}$~cm$^{-3}$. It is worth noting that this is a very large radius, likely to be an order of magnitude in excess of the actual value. With these parameters, Eqn~\ref{eqn:Sigma_X} gives $\Sigma_X \leq 8.3 \times 10^{-19}{\rm\,ergs\,s^{-1}\,cm^{-2}\,arcsec^{-2}}$. For comparison, the diffuse emission detected in the inner filament, but slightly further away from the site of star-formation \citep{2004ApJ...617..209E} has a flux $S_X \gtrsim 6 \times 10^{-15}$~ergs\,s$^{-1}$\,cm$^{-2}$ and is $\sim 300$~arcsec$^2$ in size, implying a surface brightness of $\Sigma_{X{\rm ,\,obs}} \gtrsim 2 \times 10^{-17}$~ergs\,s$^{-1}$\,cm$^{-2}$\,arcsec$^{-2}$. This is a $2.5\sigma$ {\it Chandra} detection for an integration time of 65 kiloseconds. Since sensitivity goes as the square root of integration time, we don't expect the predicted surface brightess to be observable until the integration time is extended by a factor of 580 to 38 megaseconds.  We note that the maximum enhancement of X-ray emissivity due to the bow shock is around an order of magnitude \citep{2002MNRAS.335..610A}, and therefore the suggested weak shock would be undetectable in available X-ray observations. This is confirmed by simulations. \citet{2007ApJ...665.1129K} report X-ray detection of a shock around the south-west lobe of Centaurus A. Typical temperatures in our simulation around the deflected (north-east) lobe are higher, but the maximum density is lower. X-ray brightness scales as ${n_e}^2 T^{1/2}$, and the global emissivity from the gas turns out to be about a factor of five less than the measured values from \citet{2007ApJ...665.1129K}.

\subsubsection{Filament size}

If a weak (mildly supersonic) bow shock is indeed responsible for the observed NUV and \Ha\/ emission, the age of stars at the south-west tip should be similar to the shock wave travel time up the filament. The extent of the filament is $\sim 125$~arcsec, or 2.2~kpc.  Given a sound speed of $300-400$~km\,s$^{-1}$ \citep{2004ApJ...617..209E},  
and assuming a bow shock speed of $600-900$~km\,s$^{-1}$ (sec Section~\ref{sec:x-ray_emission} above), the expected age of the young stellar population is $2.4 - 3.6$~Myrs, consistent with the values derived in Section~\ref{sec:sf}.

\subsection{X-ray emission in the filament}
\label{sec:predictions}

In our picture, both the shocked and ionized gas will be uplifted. Both the \Ha\/ and X-ray surface brightness are proportional to the square of the gas density (e.g. Eqn~\ref{eqn:Sigma_X} of this paper and Eqn~7 of \citet{1993ApJ...414..510S}. Therefore, the ratio $\Sigma_{H\alpha} / \Sigma_X$ should remain roughly constant and we can use the \Ha\/ luminosities of \citet{1991MNRAS.249...91M} to predict the expected X-ray surface brightness along the filament. Diffuse X-ray emission is observed around regions B and C. The predicted values for Regions~E and F are $6.8$ and $6.4 \times 10^{-18}$~ergs\,s$^{-1}$\,cm$^{-2}$\,arcsec$^{-2}$ respectively, a factor of three lower than the $2.5\sigma$ surface brightness above the background detected by \citet{2004ApJ...617..209E}. A ten-fold increase in the exposure time should therefore just pick up the diffuse X-ray emission.

\subsection{Masses}

The X-ray and \Ha\/ surface brightness can be used to estimate the initial mass of the cloud that has been disrupted to form the filament.

\subsubsection{X-ray}

The X-ray emission is spatially diffuse, and it is likely that a substantial part of it is not detected by {\it Chandra} in the parts of the filament further away from the nucleus. Under the assumption of X-ray emission scaling with \Ha\, we use the \citet{1991MNRAS.249...91M} values to interpolate the X-ray surface brightness profile. This yields $\left( \frac{\Sigma_X}{10^{-16}{\rm\,ergs\,s^{-1}\,cm^{-2}arcsec^{-2}}} \right) = 1.04 \left( \frac{\theta_A}{\rm arcsec} \right)^{-0.666}$ where $\theta_A$ is the distance along the filament to Region~A of \citet{1991MNRAS.249...91M}. Eqn~\ref{eqn:Sigma_X} then relates this surface brightness to a gas density. Assuming a filament width of 20~arcsec \citep{2004ApJ...617..209E}, integration along the filament yields an estimate for total X-ray gas mass of $M_X = 3.7 \times 10^5$~\Msun\/. 

\subsubsection{H$\alpha$}

A similar estimate for the \Ha\/-emitting gas yields $M_{H\alpha} = 1.2 \times 10^6$~\Msun\/. However, unlike the diffuse X-ray emission, the \Ha\/-flux is known to be mostly contained in knots, and therefore this value is very much an upper limit. A better estimate is provided by adding up the contributions from the H$\alpha$ knots. We can take $n_e \propto \Sigma_{H\alpha}^{1/2}$, where the constant of proportionality is derived using Eqn~\ref{eqn:n_e} for Region~A. This yields $M_{H\alpha} = 2.1 \times 10^5$~\Msun\, a value which is much more consistent with the observed masses of nearby H\,I clouds of $\leq 8 \times 10^5$~\Msun\/ \citep{2010A&A...515A..67S}.

It should be noted that the above calculation implicitly assumes the same scaling between electron density $n_e$ and surface brightness $\Sigma_{H\alpha}$ along the H$\alpha$ filament. However, this may not be true, since we have suggested that the \Ha\/ emission in Region~A may be heavily influenced by the photoionizing flux from young stars, while it is dominated by shocks further along the filament. Using the gas density of $n_e \approx 1$~cm$^{-3}$ derived by \citet{1993ApJ...414..510S} for their shock model, and assuming the filaments to be circularly symmetric about the major axis, we derive a mass of $M_{H\alpha} = 1.5 \times 10^5$~\Msun\/, in good agreement with the value derived above from the H$\alpha$-flux.

\subsubsection{Cloud constituents}

There is no detectable H\,I-flux in the vicinity of the inner filament, and the mass of young stars in the south-west tip is estimated from $F814W$ photometry to be $M_{\rm stars} \approx (2-8) \times 10^3$~\Msun\/. Thus, we estimate the total pre-ablation cloud mass of $M_{\rm cloud} = M_X + M_{H\alpha} + M_{\rm stars} \approx 5.8 \times 10^5$~\Msun\/. Recently, \citet{2010A&A...515A..67S} have reported observations of two new H\,I clouds near the inner filament, of masses $\leq 8 \times 10^5$~\Msun\/. It is therefore entirely plausible that another one of these clouds could have been overrun by a weak shock within the last few Myrs.

The mass of the star-forming component is also entirely consistent with this picture. A passing shock will only ablate and efficiently heat the more diffuse gas components of a molecular cloud. Dense molecular clumps are radiative (i.e. they have very short cooling times), and as a result shock passage can only trigger star-formation by compressing the gas. Typical molecular clump masses of $>10^3$~\Msun\/ \citep{2001ApJ...549..979K, 2006MNRAS.372..457S} are consistent with the derived mass in young stars.

\section{Summary}
\label{sec:summary}

In this paper, we presented high-resolution WFC3 NUV observations of the inner filament of Centaurus~A. We found a significant population of recently formed stars at the south-west tip of the filament, and no such population further along the filament. Stellar population modeling reveals that these stars have ages of only a few Myrs.

The location and age of this stellar population suggests an explanation for the origin of the multi-wavelength structure of the inner filament. As originally proposed by \citet{1993ApJ...414..510S}, the observed optical line emission is driven by shocks originating from gas clumps. We propose that these shocks arise due to a weak bow shock propagating through the diffuse interstellar medium, from a location near the inner northern radio lobe. The shock compresses and ablates a molecular gas cloud, triggering the observed star-formation at the south-west tip. The dragging out of the diffuse cloud gas and subsequent shocks give rise to the observed optical narrow-line and X-ray emission.

We show that the derived stellar ages for the UV-bright population are consistent with the gas uplifting timescale. We predict that much deeper X-ray observations should reveal more faint, diffuse emission along the filament. More generally, this mechanism should be applicable to any galaxy or galaxy cluster containing gas clumps and capable of driving weak bow shocks. 

Positive feedback is likely to be present in many systems, perhaps even more so at high redshift where the gas supply and gas densities are much higher and interaction between AGN and gas is much more frequent. This type of process may play an important role in the rapid buildup of stellar mass in massive galaxies \citep[e.g.][]{2005MNRAS.364.1337S}. This study is therefore a useful laboratory to trace this process, which is both useful from a purely observational viewpoint but also as a potential constraint on models of galaxy formation at high redshift.

\section*{Acknowledgements}

This paper is based on Early Release Science observations made by the WFC3 Scientific Oversight Committee.  We are grateful to the Director of the Space Telescope Science Institute for awarding Director's Discretionary time for this program.  Finally, we are deeply indebted to the astronauts of STS-125 for rejuvenating HST.  Support for Program numbers 11359/60 as provided by NASA through a grant from the Space Telescope Science Institute, which is operated by the Association of Universities for Research in Astronomy, Incorporated, under NASA contract NAS5-26555.  RMC acknowledges funding from STFC through research grant DBRPDV0, and support from the James Martin Institute, Oxford through a James Martin Fellowship.  S.S. thanks the Australian Research Council and New College, Oxford for research fellowships. S.K. acknowledges a Research Fellowship from the Royal Commission for the Exhibition of 1851, an Imperial College Junior Research Fellowship, a Senior Research Fellowship form Worcester College, Oxford and support from the BIPAC institute.

\bibliography{CenA_bibtex.bib}

\newpage
\onecolumn
\appendix
\section{Table of photometry}

\begin{center}
\begin{longtable}{lllcccr}
\caption{Table of photometry.  Includes only objects detected in both $F225W$ and at least one other filter.  Photometry of objects 12, 15 and 16 was performed using small apertures, with aperture corrections derived using {\sc ishape} \citep{1999A&AS..139..393L}.  Visual inspection of objects 25 and 29 revealed that the $F814W$ detections were not coincident with those in $F225W$ and $F606W$.  Photometry of the globular cluster candidates, objects 81 and 109 was performed using large apertures with radii of 3$\times$ each object's FWHM in $F606W$ -- 0.80$\arcsec$ and 0.47$\arcsec$ respectively.  
}
\label{tab:phot_table} \\

\hline\hline
\multicolumn{1}{l}{I.D.} & 
\multicolumn{1}{l}{R.A.\footnotemark[1]} & 
\multicolumn{1}{l}{Decl.\footnotemark[1]} & 
\multicolumn{1}{c}{F225W} & 
\multicolumn{1}{c}{F606W} & 
\multicolumn{1}{c}{F814W} &  
\multicolumn{1}{c}{Comment}\\
\hline
\endfirsthead

\multicolumn{7}{c}{\tablename\ \thetable{} -- continued from previous page} \\
\hline
\multicolumn{1}{l}{I.D.} & 
\multicolumn{1}{l}{R.A.\footnotemark[1]} & 
\multicolumn{1}{l}{Decl.\footnotemark[1]} & 
\multicolumn{1}{c}{F225W} & 
\multicolumn{1}{c}{F606W} & 
\multicolumn{1}{c}{F814W} & 
\multicolumn{1}{c}{Comment}\\
\hline
\endhead

\hline
\multicolumn{4}{l}{$^1$Coordinates measured in WFC3 F225W data} & \multicolumn{3}{r}{Continued on next page}
\endfoot

\hline
\multicolumn{7}{l}{$^1$Coordinates measured in WFC3 F225W data}
\endlastfoot

\multicolumn{7}{l}{{\bf Southwest tip - Region 1}}\\
   1 & 13:26:03.253 & -42:57:19.36 &  23.55(0.11) &  24.68(0.06) &   24.62(0.09) &\\
   2 & 13:26:03.264 & -42:57:19.18 &  22.29(0.05) &  24.33(0.04) &   24.21(0.06) &\\
   3 & 13:26:03.304 & -42:57:19.04 &  21.16(0.05) &  23.44(0.05) &   23.60(0.04) &\\
   4 & 13:26:03.274 & -42:57:18.88 &  22.01(0.07) &  24.28(0.05) &   24.35(0.08) &\\
   5 & 13:26:03.282 & -42:57:18.94 &  22.84(0.11) &  24.61(0.05) &   25.00(0.14) &\\
   6 & 13:26:03.297 & -42:57:18.58 &  22.57(0.08) &  24.42(0.12) &   24.77(0.06) &\\
   7 & 13:26:03.328 & -42:57:18.38 &  23.45(0.08) &  25.70(0.15) &   25.61(0.15) &\\
   8 & 13:26:03.320 & -42:57:18.36 &  22.91(0.06) &  24.83(0.05) &   24.87(0.09) &\\
   9 & 13:26:03.264 & -42:57:18.34 &  24.23(0.11) &  25.87(0.10) &   26.50(0.27) &\\
  10 & 13:26:03.207 & -42:57:18.31 &  22.31(0.05) &  24.65(0.10) &   24.81(0.09) &\\
  11 & 13:26:03.325 & -42:57:18.23 &  22.27(0.04) &  24.39(0.05) &   24.53(0.07) &\\
  12 & 13:26:03.418 & -42:57:18.04 &  21.04(0.03) &  22.46(0.02) &   22.38(0.03) & Extended object\\
  13 & 13:26:03.372 & -42:57:17.82 &  24.12(0.11) &  24.75(0.11) &   25.09(0.10) &\\
  14 & 13:26:03.502 & -42:57:17.77 &  22.44(0.06) &  24.43(0.04) &   24.68(0.09) &\\
  15 & 13:26:03.424 & -42:57:17.63 &  20.32(0.01) &  21.45(0.01) &   21.25(0.02) & Extended object\\
  16 & 13:26:03.427 & -42:57:17.49 &  20.78(0.02) &  21.14(0.01) &   21.13(0.02) & Extended object\\
  17 & 13:26:03.453 & -42:57:17.36 &  23.69(0.11) &  23.72(0.09) &   23.71(0.09) &\\
  18 & 13:26:03.408 & -42:57:17.38 &  22.98(0.11) &  23.39(0.10) &   23.80(0.17) &\\
  19 & 13:26:03.375 & -42:57:17.34 &  23.20(0.06) &  24.28(0.05) &   24.24(0.05) &\\
  20 & 13:26:03.332 & -42:57:17.28 &  24.16(0.12) &  24.61(0.07) &   24.73(0.09) &\\
\\
\multicolumn{7}{l}{{\bf Southwest tip - Region 2}}\\
  21 & 13:26:02.940 & -42:57:21.80 &  23.27(0.06) &  24.64(0.07) &   24.50(0.07) &\\
  22 & 13:26:02.968 & -42:57:21.54 &  23.61(0.09) &  25.65(0.10) &   25.75(0.20) &\\
  23 & 13:26:02.856 & -42:57:21.09 &  24.12(0.06) &  26.02(0.05) &   26.00(0.12) &\\
  24 & 13:26:02.904 & -42:57:20.80 &  23.99(0.08) &  25.99(0.07) &   27.27(0.50) & Very weak F814W detection\\
  25 & 13:26:02.856 & -42:57:20.73 &  24.29(0.10) &  25.38(0.07) &   24.73(0.05) & F814W detection is different object\\
  26 & 13:26:02.887 & -42:57:20.46 &  20.73(0.04) &  23.03(0.06) &   23.09(0.03) &\\
  27 & 13:26:02.889 & -42:57:19.93 &  24.28(0.13) &  26.28(0.09) &   26.22(0.23) &\\
  28 & 13:26:02.878 & -42:57:19.73 &  21.14(0.04) &  23.53(0.06) &   23.64(0.03) &\\
  29 & 13:26:02.877 & -42:57:19.41 &  24.72(0.17) &  25.71(0.11) &   24.04(0.05) & F814W detection is different object\\
\\
\multicolumn{7}{l}{{\bf Rest of WFC3/ACS field}}\\
  30 & 13:26:03.538 & -42:58:38.93 &  24.36(0.12) &   -- &   23.64(0.03) & Outside F606W FOV\\
  31 & 13:26:07.384 & -42:58:07.24 &  24.54(0.10) &   -- &   24.72(0.09) & Outside F606W FOV\\
  32 & 13:26:03.736 & -42:58:02.39 &  23.03(0.08) &   -- &   -- & Saturated/non-linear F606W \& F814W\\
  33 & 13:26:09.251 & -42:57:58.56 &  23.42(0.07) &   -- &   18.83(0.07) & Outside F606W FOV\\
  34 & 13:26:01.558 & -42:57:46.75 &  21.98(0.04) &   -- &   16.91(0.04) & Saturated/non-linear F606W\\
  35 & 13:26:00.483 & -42:57:46.39 &  23.51(0.09) &  24.98(0.05) &   24.95(0.08) & \\
  36 & 13:26:00.606 & -42:57:44.40 &  24.15(0.11) &  25.46(0.05) &   25.46(0.16) &\\
  37 & 13:26:13.163 & -42:57:37.06 &  23.50(0.09) &   -- &	20.92(0.04) & Outside F606W FOV\\
  38 & 13:26:01.513 & -42:57:35.51 &  24.69(0.13) &  19.76(0.06) &   18.97(0.03) &\\
  39 & 13:26:01.862 & -42:57:33.35 &  22.27(0.04) &   -- &   16.60(0.02) & Saturated/non-linear F606W\\
  40 & 13:26:15.019 & -42:57:28.38 &  24.29(0.08) &   -- &   25.09(0.09) & Outside F606W FOV\\
  41 & 13:26:02.694 & -42:57:28.30 &  23.74(0.09) &  19.94(0.07) &   19.22(0.02) &\\
  42 & 13:26:06.972 & -42:57:23.87 &  16.85(0.02) &   -- &   -- & Saturated/non-linear F606W \& F814W\\
  43 & 13:26:01.164 & -42:57:23.88 &  24.27(0.09) &  25.25(0.09) &   24.74(0.06) &\\
  44 & 13:26:06.314 & -42:57:23.22 &  22.79(0.05) &  24.83(0.06) &   24.78(0.07) &\\
  45 & 13:26:01.534 & -42:57:22.67 &  23.91(0.09) &  25.36(0.10) &   25.19(0.10) &\\
  46 & 13:26:01.636 & -42:57:21.04 &  23.81(0.12) &  25.17(0.06) &   25.36(0.10) &\\
  47 & 13:26:09.319 & -42:57:19.63 &  23.00(0.08) &   -- &   20.52(0.02) & Outside F606W FOV\\
  48 & 13:26:04.809 & -42:57:16.76 &  23.64(0.06) &  25.60(0.09) &   25.62(0.14) &\\
  49 & 13:26:04.841 & -42:57:16.05 &  22.03(0.05) &  23.95(0.11) &   24.18(0.08) &\\
  50 & 13:26:11.406 & -42:57:15.16 &  24.69(0.11) &   -- &   18.31(0.03) & Outside F606W FOV\\
  51 & 13:26:14.150 & -42:57:13.46 &  18.67(0.04) &   -- &   -- & Outside F606W FOV; saturated F814W\\
  52 & 13:26:05.846 & -42:57:11.29 &  20.56(0.03) &  19.40(0.07) &   18.65(0.02) &\\
  53 & 13:26:03.353 & -42:57:10.85 &  21.60(0.04) &   -- &   17.61(0.02) & Saturated/non-linear F606W\\
  54 & 13:26:11.240 & -42:57:07.72 &  24.26(0.08) &   -- &   23.14(0.05) & Outside F606W FOV\\
  55 & 13:26:08.780 & -42:57:07.60 &  24.67(0.12) &   -- &   21.34(0.03) & Outside F606W FOV\\
  56 & 13:25:58.130 & -42:57:06.47 &  23.81(0.09) &   -- &   17.49(0.02) & Saturated/non-linear F606W\\
  57 & 13:26:00.599 & -42:57:03.44 &  23.05(0.06) &  23.71(0.07) &   23.56(0.07) &\\
  58 & 13:26:02.147 & -42:57:03.33 &  23.23(0.06) &  20.61(0.08) &   19.91(0.02) &\\
  59 & 13:26:02.016 & -42:57:02.87 &  24.17(0.09) &  25.46(0.09) &   24.93(0.10) &\\
  60 & 13:25:59.413 & -42:57:02.50 &  24.37(0.14) &  24.52(0.06) &   24.43(0.06) &\\
  61 & 13:25:59.159 & -42:57:01.92 &  23.93(0.08) &  25.02(0.05) &   24.90(0.08) &\\
  62 & 13:25:59.050 & -42:56:58.89 &  23.23(0.07) &  24.50(0.07) &   24.43(0.05) &\\
  63 & 13:25:59.644 & -42:56:58.28 &  23.57(0.05) &  24.60(0.05) &   24.74(0.08) &\\
  64 & 13:25:57.687 & -42:56:56.84 &  23.67(0.05) &  23.15(0.06) &   23.02(0.03) &\\
  65 & 13:25:57.128 & -42:56:56.52 &  22.73(0.05) &  21.01(0.09) &   20.33(0.04) &\\
  66 & 13:26:00.489 & -42:56:55.08 &  24.37(0.11) &  25.36(0.09) &   25.61(0.16) &\\
  67 & 13:26:12.125 & -42:56:54.80 &  20.47(0.04) &   -- &   17.16(0.02) & Outside F606W FOV\\
  68 & 13:25:59.748 & -42:56:53.16 &  24.01(0.07) &  25.18(0.09) &   25.14(0.11) &\\
  69 & 13:26:10.273 & -42:56:51.09 &  21.95(0.03) &   -- &   17.90(0.02) & Outside F606W FOV\\
  70 & 13:26:00.028 & -42:56:51.07 &  24.84(0.17) &  26.38(0.08) &   25.57(0.12) &\\
  71 & 13:26:05.809 & -42:56:50.23 &  20.69(0.03) &   -- &   16.38(0.05) & Saturated/non-linear F606W\\
  72 & 13:26:00.079 & -42:56:49.87 &  23.65(0.06) &  24.97(0.08) &   24.94(0.07) &\\
  73 & 13:26:00.573 & -42:56:49.71 &  23.36(0.06) &  23.08(0.06) &   22.88(0.03) &\\
  74 & 13:26:08.224 & -42:56:49.50 &  23.94(0.11) &  24.13(0.08) &   23.69(0.06) &\\
  75 & 13:26:03.122 & -42:56:48.97 &  22.01(0.04) &   -- &   16.82(0.02) & Saturated/non-linear F606W\\
  76 & 13:25:59.577 & -42:56:45.09 &  23.73(0.06) &  21.38(0.06) &   20.59(0.03) &\\
  77 & 13:25:58.771 & -42:56:40.10 &  24.19(0.09) &  25.09(0.03) &   24.69(0.08) &\\
  78 & 13:25:56.932 & -42:56:39.31 &  23.23(0.07) &   -- &   16.98(0.02) & Saturated/non-linear F606W\\
  79 & 13:26:12.608 & -42:56:34.21 &  24.44(0.21) &   -- &   25.49(0.11) & Outside F606W FOV\\
  80 & 13:26:06.155 & -42:56:33.57 &  21.77(0.04) &  23.77(0.06) &   23.89(0.03) &\\
  81 & 13:26:05.386 & -42:56:32.28 &  19.87(0.01) &  17.17(0.01) &   16.23(0.01) & Globular cluster; some non-linear pixels F606W\\
  82 & 13:25:57.294 & -42:56:30.27 &  23.96(0.06) &  20.10(0.06) &   19.36(0.02) &\\
  83 & 13:26:06.849 & -42:56:28.96 &  23.99(0.09) &  25.26(0.04) &   24.84(0.09) &\\
  84 & 13:25:59.533 & -42:56:27.51 &  23.16(0.05) &  24.45(0.09) &   24.09(0.06) &\\
  85 & 13:25:59.684 & -42:56:25.06 &  23.79(0.08) &  25.37(0.10) &   25.28(0.11) &\\
  86 & 13:25:59.413 & -42:56:25.06 &  23.23(0.05) &  25.08(0.08) &   25.13(0.09) &\\
  87 & 13:25:59.310 & -42:56:24.91 &  20.94(0.03) &  22.65(0.05) &   22.65(0.03) &\\
  88 & 13:26:03.337 & -42:56:24.07 &  19.84(0.04) &   -- &   -- & Saturated/non-linear F606W \& F814W\\
  89 & 13:25:59.278 & -42:56:23.75 &  21.48(0.03) &  23.50(0.03) &   23.44(0.03) &\\
  90 & 13:25:59.597 & -42:56:23.20 &  23.59(0.07) &  25.32(0.07) &   25.56(0.15) &\\
  91 & 13:25:59.385 & -42:56:22.63 &  23.14(0.06) &  25.04(0.06) &   24.98(0.07) &\\
  92 & 13:25:59.546 & -42:56:22.42 &  24.36(0.11) &  26.17(0.05) &   26.06(0.16) &\\
  93 & 13:25:59.388 & -42:56:23.26 &  24.78(0.11) &  25.96(0.07) &   25.85(0.20) &\\
  94 & 13:26:08.344 & -42:56:22.03 &  21.99(0.03) &   -- &   17.39(0.04) & Saturated/non-linear F606W\\
  95 & 13:25:59.699 & -42:56:21.36 &  21.72(0.04) &  23.93(0.05) &   23.99(0.05) &\\
  96 & 13:25:59.297 & -42:56:21.17 &  21.72(0.04) &  23.76(0.05) &   23.64(0.04) &\\
  97 & 13:25:59.357 & -42:56:20.67 &  22.33(0.04) &  24.47(0.06) &   24.64(0.09) &\\
  98 & 13:25:58.745 & -42:56:20.31 &  23.77(0.07) &  25.30(0.05) &   24.91(0.09) &\\
  99 & 13:26:05.651 & -42:56:19.00 &  23.03(0.06) &  19.57(0.07) &   18.60(0.03) &\\
 100 & 13:26:09.942 & -42:56:14.55 &  24.07(0.09) &   -- &   18.38(0.02) & Saturated/non-linear F606W\\
 101 & 13:26:05.969 & -42:56:11.44 &  20.77(0.03) &   -- &   -- & Saturated/non-linear F606W \& F814W\\
 102 & 13:26:07.718 & -42:56:10.24 &  22.74(0.07) &  22.60(0.11) &   22.45(0.04) &\\
 103 & 13:26:03.636 & -42:56:09.33 &  21.10(0.04) &   -- &   -- & Saturated/non-linear F606W \& F814W\\
 104 & 13:26:05.352 & -42:56:00.29 &  24.43(0.09) &  20.03(0.08) &   18.96(0.03) &\\
 105 & 13:26:05.220 & -42:55:58.91 &  20.30(0.04) &   -- &   -- & Saturated/non-linear F606W \& F814W\\
 106 & 13:26:10.200 & -42:55:58.07 &  22.74(0.04) &  19.65(0.08) &   18.88(0.02) &\\
 107 & 13:26:06.602 & -42:55:56.69 &  23.23(0.05) &   -- &   17.81(0.02) & Saturated/non-linear F606W\\
 108 & 13:26:10.065 & -42:55:55.60 &  21.41(0.03) &   -- &   17.95(0.04) & Saturated/non-linear F606W\\
 109 & 13:26:04.165 & -42:55:44.60 &  21.32(0.02) &  19.04(0.01) &   18.22(0.01) & Globular cluster\\
 110 & 13:26:07.989 & -42:55:44.09 &  21.25(0.04) &   -- &   -- & Saturated/non-linear F606W \& F814W\\
 111 & 13:26:06.737 & -42:55:34.92 &  24.67(0.20) &  27.21(0.11) &   25.90(0.18) &\\
\end{longtable}
\end{center}

\label{lastpage}
\end{document}